# An open source massively parallel solver for Richards equation: mechanistic modelling of water fluxes at the watershed scale

L. Orgogozo[1*], N. Renon[2], C. Soulaine[3], F. Hénon[3], S. K. Tomer[4], D. Labat[1], O. S. Pokrovsky[1,5], M. Sekhar[6], R. Ababou[3], M. Quintard[3,7]

[1] GET (Géosciences Environnement Toulouse), Observatoire Midi-Pyrénées, Université Paul Sabatier, Université de Toulouse; 14 avenue Édouard Belin, 31400 Toulouse, France
[2] CALMIP Toulouse University Computing Center/DTSI, Université Paul Sabatier, Université de Toulouse; 118 route de Narbonne, 31400 Toulouse, France
[3] Université de Toulouse ; INPT, UPS ; IMFT (Institut de Mécanique des Fluides de Toulouse); Allée Camille Soula, F-31400 Toulouse, France
[4] CESBIO (Centre d'Etude Spatiale de la BIOsphère), Observatoire Midi-Pyrénées, Université Paul Sabatier, Université de Toulouse; 18 avenue Edouard Belin, BPI 2801, 31401 Toulouse cedex 9, France
[5] BIO-GEO-CLIM Laboratory, Tomsk State University; Tomsk, Russia
[6] Indian Institute of Science, Department of Civil Engineering, Indo-French cell for Water Sciences; Bangalore – 560 012
[7] CNRS; IMFT; F-31400 Toulouse, France
* Corresponding author :E-mail: laurent.orgogozo@get.obs-mip.fr, tel : +33 61 33 25 74

## Abstract

In this paper we present a massively parallel open source solver for Richards equation, named the RichardsFOAM solver. This solver has been developed in the framework of the open source generalist computational fluid dynamics tool box OpenFOAM® and is capable to deal with large scale problems in both space and time. The source code for RichardsFOAM may be downloaded from the CPC program library website.

It exhibits good parallel performances (up to ~90% parallel efficiency with 1024 processors both in strong and weak scaling), and the conditions required for obtaining such performances are analysed and discussed. These performances enable the mechanistic modelling of water fluxes at the scale of experimental watersheds (up to few square kilometres of surface area),, and on time scales of decades to a century. Such a solver can be useful in various applications, such as environmental engineering for long term transport of pollutants in soils, water engineering for assessing the impact of land settlement on water resources, or in the study of weathering processes on the watersheds.

**Programme summary**

Manuscript Title: An open source massively parallel solver for Richards equation: Mechanistic modelling of water fluxes at the watershed scale

Authors: Laurent Orgogozo, Nicolas Renon, Cyprien Soulaine, Florent Hénon, Sat Kumar Tomer, David Labat, Oleg S. Pokrovsky, Muddu Sekhar, Rachid Ababou, Michel Quintard

Program title: RichardsFOAM

Catalogue identifier: -

Distribution format: tar.gz

Journal reference: Comput. Phys. Commun., submitted

Programming language: C++

Computer: any x86, tested only on 64-bit machines.

Operating system: Generic Linux.

Classification: 13

External routines: OpenFOAM® (version 2.0.1 or later)

Nature of problem: This software solves the non-linear three-dimensional transient Richards equation, which is a very popular model for water transfer in variably saturated porous media (e.g.: soils). It is designed to take advantage of the massively parallel computing performance of OpenFOAM®. The goal is to be able to model natural hydrosystems on large temporal and spatial scales.

Solution method: A mixed implicit (FVM in the object oriented OpenFOAM® framework) and explicit (FVC in the object oriented OpenFOAM® framework) discretization of the equation with a backward time scheme is coupled with a linearization method (Picard algorithm). Due to the linearization loop the final solution of each time step tends towards a fully implicit solution. The implementation has been carried out with a concern for robustness and parallel efficiency.

Restrictions: the choice of the maximum and initial time steps must be made carefully in order to avoid stability problems. A careful convergence study of mesh cell size, linear solver precision and linearization method precision must be undertaken for each considered problem, depending on the precision required for the expected results, the spatial and temporal scales at stake, and so on. Finally, the solver in its current version only handles meshes with a constant cell volume (a crash will not necessarily occur with an irregular mesh but some problems may arise with the convergence criterion of the linearization method).

Running time: highly variable, depending on the mesh size and the number and nature of cores involved. The test run provided requires less than 2 seconds on a 64 bit machine with Intel®CoreTMi7-2760QM CPU @ 2.40GHz x8 and 3.8 Gigabytes of RAM.

# 1   Introduction

Many applications in the geosciences involve transfers of water in variably saturated porous media, such as soils. From the point of view of water engineering, the direct infiltration of rainwater into soils is the main recharge of aquifers, and is also their main source of pollution, for instance through infiltration of dissolved nitrates and pesticides. Accurate modelling of water transfer in soils is thus important for water engineering applications. On the other hand, water transfer in porous media controls water content in the soil profile, which is one of the driving parameters of weathering [1], the key process in the carbon cycle [2].

It is generally admitted (for example, [3-5]) that mechanistic approaches are the "gold standard" [5] for modelling natural water systems, since they allow quantitative and predictive assessment of water flows. Such tools are thus of great interest for studying the impact of global changes on weathering processes, which requires modelling under evolving climatic conditions (e.g.: "dry" or "wet" warming). The design of water management infrastructures (e.g.: dams, wells) also leads to the necessity of modelling the water transfers within changing water systems, and thus requires the use of mechanistic modelling approaches. In order to perform mechanistic modelling of water flows in soils, it is necessary to be able to quantitatively assess the spatial and temporal evolution of water pressure and water content. The most widely used approach in order to reach this goal is numerical resolution of the three dimensional Richards equation [6]. This equation is an approximate solution of the general two-phase flow in porous media model based on generalized Darcy's laws [7], assuming that the pressure gradient in the gas phase is small. This assumption may be questionable under certain circumstances, in particular trapped gas phase in less permeable lenses. Similarly, the generalized Darcy's law model itself is not entirely supported by upscaling theories ([8-13]; etc) and is the subject of current research. However, up to now, Richards equation is still the model of choice in engineering practice and provides applicable results in many cases. In addition, Richards equation is used in applications other than the modelling of natural water system, such as nuclear waste repository (e.g.: [14]) or proton exchange membrane fuel cell (e.g.: [15]).

Richards equation is based on Darcy's law with pressure-dependent, or moisture dependent, hydraulic conductivity. It is applicable under the usual conditions of pressure and temperature, with water considered an incompressible fluid, with the air phase remaining connected to the atmosphere (fixed air pressure) and having a negligible viscosity. Additionally, the hysteresis effects that may be encountered in the case of successive imbibition/drainage cycles [17] are neglected in this work. The Richards approach then leads to the following governing equation for water flow in a variably saturated porous medium:

$$\frac{\partial \theta}{\partial t} = C(h)\frac{\partial h}{\partial t} = \nabla \cdot \left( \mathrm{K}(h) \cdot \nabla (h+z) \right) \qquad (1)$$

In this equation, $h$ is the pressure head expressed as length of water column [m], $z$ is the vertical coordinate [m] (oriented upward), K($h$) is the hydraulic conductivity of the unsaturated porous medium [m.s$^{-1}$], $\theta(h)$ is the volumetric water content [m$^3$.m$^{-3}$] and C($h$) is the capillary capacity (also called specific moisture capacity) of the unsaturated porous medium [m$^{-1}$]. Leaving $\partial \theta / \partial t$ on the left hand side leads to the so-called mixed formulation of the Richards equation in numerical implementations. With $C(h)\partial h/\partial t$ on the left hand side, we have the pressure formulation of Richards equation, which is that adopted in this work, and is also historically the original equation of Richards [6]. Water content formulations also exist, but they are limited to strictly unsaturated situations without the possibility of pressure build up in saturated zones. One can refer to [17] for a detailed discussion of the mixed formulation for implicit 3D finite volumes, and to [18] for a systematic comparison of these various formulations for 1D unsaturated flows.

The main complexity of the Richards equation lies in the non-linearities due to the pressure dependent hydraulic conductivity $K(h)$ and of the pressure dependent capillary capacity $C(h)$. These dependencies, as well as the water content – pressure relationship (the water retention curve $\theta(h)$), are treated with empirical models such as that of Brooks and Corey [19] or the van Genuchten model [20] which is based in part on the functional model of Mualem [21]. The van Genuchten/Mualem model is very popular and will be the one used in this work, but adopting a different model would not pose any implementation difficulty. Here is a brief summary of the van Genuchten model, expressed in terms of $\theta(h)$, $C(h)$ and $K(h)$:

Retention curve $\theta = f(h)$ $[-]$

$$\begin{cases} \theta = \theta_s \text{ if } h \geq 0 \\ \theta = (\theta_s - \theta_r)\left(1 + (-\alpha h)^n\right)^{-(1-(1/n))} + \theta_r \text{ if } h < 0 \end{cases} \quad (2)$$

Capillary capacity $C(h)$ $\left[L^{-1}\right]$,

$$\begin{cases} C(h) = S \text{ if } h \geq 0 \\ C(h) = (\theta_s - \theta_r)\alpha n\left(1 - (1/n)\right)(-\alpha h)^{n-1}\left(1 + (-\alpha h)^n\right)^{-(2-(1/n))} \text{ if } h < 0 \end{cases} \quad (3)$$

Hydraulic conductivity $\left[L.T^{-1}\right]$,

$$\begin{cases} K(h) = K_s \text{ if } h \geq 0 \\ K(h) = K_s\left(\left(1+(-\alpha h)^n\right)^{-(1-(1/n))}\right)^{(1/2)}\left(1 - \left(1 - \left(\left(1+(-\alpha h)^n\right)^{-(1-(1/n))}\right)^{(n/n-1)}\right)^{1-(1/n)}\right)^2 \text{ if } h < 0 \end{cases} \quad (4)$$

In addition to previously defined variables, $\theta_s$ is the saturation water content $[-]$, $\theta_r$ is the residual water content $[-]$, $K_s$ is the saturated hydraulic conductivity $\left[\text{m.s}^{-1}\right]$, $S$ is the specific storativity coefficient of the porous medium under positive pressure only $\left[\text{m}^{-1}\right]$, and finally, $\alpha$ $\left[\text{m}^{-1}\right]$ and $n$ $[-]$ are empirical parameters which are characteristics of the considered soil (see for example [22]). The inverse of $\alpha$ may be considered a characteristic capillary length scale of the medium. Therefore, the Richards equation is a non-linear diffusion/conduction PDE which contains an additional hyperbolic (gravitational) term with strongly non-linear coefficients which may be spatially variable.

Mechanistic flow modelling with the Richards equation consists of numerical resolution of this equation in the domain under consideration. In order to apply this approach to the weathering process, large scale modelling is necessary, both from the spatial and temporal point of view. From the spatial point of view, watershed scale (> several km$^2$) must be considered in order to calculate mass balances involving, for instance, weathering fluxes. From the temporal point of view, one should, for instance, consider the time scale of the observed global warming, i.e., the century scale. Such large scales may be encountered in a number of environmentally-relevant applications, such as the evolution of water resources in the context of climate change or the quantification of long term migrations of pollutants within soils. These space-time requirements constitute a major challenge for mechanistic modelling, with very long computational times and very large memory requirements to be expected [23]. As such, the application of a mechanistic modelling approach to such large space and time scales requires the use of state of the art numerical techniques and hardware, and, according to Miller et al. [5], "the efficiency of serial computers has approached its limit, and increased hardware efficiency will be primarily based on parallel computing".

A number of numerical tools for modelling of water transfers in soils are already available (e.g.: [24-33]). However, these tools do not allow (or were not tested) for the use of parallel computations. Efforts to build and test parallel solvers were then made (for instance, [4], [17], [34-40]), but these solvers have low or moderate parallel efficiencies, particularly when used in massively parallel computations (with several hundreds of cores – note that only a few were even tested with more than a hundred cores). The parallel solver for Richards equation proposed by Hardelauf et al., 2007 [41] exhibits good parallel performances up to 256 cores, but no scaling for a higher number of cores is shown. Indeed, massively parallel computations are required to model unsaturated flow processes on large spatial and temporal scales, and recent works propose tools which allow such massively parallel computations with good parallel performances ([42-47]). For example Maxwell [47] obtains, in a weak scaling exercise, parallel efficiencies of between 80% and 90% with 16 384 processors, depending on the type of preconditioner used. These are good parallel performances, and will probably be assessed depending on several factors such as the frequency of the I/Os. The originality of the present work is to propose a fully open-source solver with parallel performances on a level comparable with the current state of the art works as cited above (e.g.: [47]) within a generalist open source platform of computational fluid dynamics: OpenFOAM®. One of the advantages of using such a multiphysics toolbox is the possibility of adding easily other transport mechanisms, such as transport of chemical species, heat transfer, etc., without complex implementation problems. In addition, one can use in conjunction with RichardsFOAM the various numerical tools for pre-processing, meshing, post-processing, etc, that are (and have been) developed by the broad community of developers and users of OpenFOAM®.

In this work we propose a massively parallel 3D solver for Richards equation, RichardsFOAM, developed in the framework of OpenFOAM® ([48-50]), an open source finite volumes tool box designed to allow the use of massively parallel computing. Finite volumes schemes (structured or not) have proven to be suitable for the resolution of Richards equation (see for instance [51], [52], and some of the works cited above) and, therefore, they are well adapted to our problem. OpenFOAM® is based on an object-oriented C++ coding approach, built since the beginning of its development for massively parallel computations, and within which a large variety of numerical tools are available (meshing tools, schemes of discretization, pre-processing tools, preconditioners, linear solvers, post-processing tools, visualisation tools, etc.). It allows fast implementation of new solvers for specific equations through high-level C++ statements. Because the sources are completely open, it allows the collaborative development of new numerical tools by large groups of developers. The fact that OpenFOAM® is a generalist platform is also interesting as many tools and utilities have been developed and released by the whole community of developers and users from various scientific and engineering areas (transportation, chemical processing, marine, energy, medical systems…) on a regular basis for years. These tools and utilities may thus be used directly in conjunction with RichardsFOAM (or any other numerical tool developed within OpenFOAM®). Thus OpenFOAM allows a fast circulation of ideas and implementation innovations between the various fields of computational continuum mechanics. Some applications of OpenFOAM® to porous media flow have been recently developed (e.g.: [53]), and the potential of OpenFOAM® in the field of water sciences is more and more recognised (for instance, [54-56], [5]). Very few works present applications of OpenFOAM® in this scientific area, however (with the notable exception of Furbish and Schmeeckle [57] in the field of sediment transport). Moreover, to our knowledge, no published work focus on the application, evaluation and characterisation of the massively parallel capabilities of OpenFOAM® in the geosciences, although this has been handled in other fields such as nano-scale flows [58]. Finally, this solver for water flows in soils is intended for being the core of a more complete tool for modelling transport in continental surfaces, including thermal transfers with phase changes, reactive solute transfers and the coupling of surface flows and subsurface flows. Our main goal is to propose a fully open source, massively parallel, modelling environment for the hydrogeosciences and the related engineering applications in the framework of a widely used generalist computational fluid dynamics platform: OpenFOAM®.

The remainder of this paper is organised as follows. After a presentation of the theoretical and numerical choices made for the developed solver (the so-called RichardsFOAM solver), and a description of the hardware used, we exhibit a number of application cases: an analytical validation (in electronic supplementary material, Appendix A), a code-to-code validation and an application to a set of field data involving a heterogeneous soil column in a monsoon climate. An assessment of the parallel performance of the RichardsFOAM solver is then proposed, through two types of scaling analyses: a strong scaling analysis, and a weak scaling analysis. Finally, we discuss the potential for applying this parallel solver to the study of weathering processes at the watershed scale, which is the application for which we have handle the present work, and associated perspectives.

## 2 Materiel and methods

### 2.1 Computational issues

The implementation of RichardsFOAM was conducted while bearing in mind three major constraints closely related to the purpose of this tool: the modelling of flow in soils for large space and time scales for catchment hydrology.

The first constraint is that the memory (RAM) requested by the computation needs to be as small as possible, in order to deal with the largest possible space scales with adequate mesh refinement.

The second constraint is that the solver needs to be as stable as possible, under the broadest range of conditions, either numerical (time steps) or physical (e.g.: gravitational advection versus capillary diffusion effects). It should also be robust enough to deal with the dramatically changing hydrological regimes that can be encountered in catchment hydrology.

The third constraint is that the computation time has to be small enough to allow simulations on large time scales, up to a century, for example, in the context studying the effects of the global changes on the functioning of the critical zone (see for example [1]) or for the long term study of polluted areas (for instance [47]).

### 2.2 Implementation

We have chosen to use a simple Picard algorithm to linearize Richards equation, so as to meet the first and second constraints. Indeed, the Picard algorithm exhibits slower convergence rates than the Newton algorithm in general. However, the Newton algorithm implies the computation and storage of the Jacobian matrix associated with the problem considered, which implies a huge increase in memory needs compared to the Picard Algorithm (see for example [59-61]). A Jacobian-free Newton algorithm does exist, but it still requires more memory than the Picard algorithm. In addition, by sharply changing regimes when large numbers of linearization iterations are encountered, it can lead to higher memory needs than classical Newton linearization [62]. On the other hand, the Picard algorithm is recognised as more robust than the Newton algorithm (e.g.: [61], [63], [64]). For these two reasons, despite the fact that published massively parallel solvers for groundwater transfers do use the Newton algorithm for linearization purposes (for instance, [43], [47]), we preferred to use a Picard algorithm for our solver. As well as the classical Picard linearization algorithm, a modified Picard linearization algorithm has been implemented in order to seek mass conservation (e.g.: [59], [65]). Indeed, mass conservation problems may occur when using a pressure formulation of the Richards equation inappropriately, as mentioned (among others) by Celia et al., 1990 [66]. We have chosen here to use a classical Picard algorithm in conjunction with a chord slope approximation, which also ensures mass conservation (see the paragraph below which concerns the discretization of the time derivative). A detailed discussion about different methods of dealing with the non-linearity of Richards equation may be found in [67]. As stated previously we have used the classical van Genuchten / Mualem model ([20], [21]) to establish the non-linear dependency of hydraulic conductivity on water pressure and the water retention curve. The transition between saturated and unsaturated conditions, i.e., positive or negative water pressure, impacts strongly this dependency (see equation (3) and (4)). In order to deal with this transition, the optimised high-level functions applicable on 3D scalar fields which are available in OpenFOAM® (such as the 'sgn' function) have been used. In this way the variations of hydraulic properties of the porous medium with the water pressure may be handled using only basic operations (multiplications and additions) on scalar fields, which results in a highly scalable implementation. One can see in Appendix B (in electronic supplementary material) an illustration of the good scaling behaviour of this implementation in case of the occurrence of a transition between saturated and unsaturated conditions.

For solving the linear systems, we chose a Krylov linear solver with a diagonal incomplete Cholesky preconditioner (DIC-PCG). A Preconditioned Conjugate Gradient is a classical approach for solving linear systems resulting in the linearization of Richards equation (for example, [68-70], [39], [37], [47]). One can find an extensive discussion on the preconditioning methods for the parallel resolution of Richards equation with a Picard linearization and a Krylov linear solver in Herbst et al., 2008 [37]. They found that the best classes of methods among those tested were the incomplete Cholesky and the multigrid methods, with slightly more efficiency with the tested multigrid method (algebraic multigrid). They made their tests with a small number of cores (32 and 64), however, and it is well known that multigrid methods lead to load balance problems when used with a large number of cores (i.e., hundreds or thousands). This is the reason we chose an incomplete Cholesky preconditioning method rather than a multigrid preconditioning method. Regarding the linear solver itself, it has been shown that Conjugate Gradient solvers are particularly suitable for parallel computing, because only nonzero elements need to be stored in this method [71]. One can note that more efficient linear solvers than Krylov solvers (at least for serial computations or parallel computations with low numbers of cores) have been implemented in OpenFOAM®, for example the geometric agglomerated algebraic multigrid linear solver (GAMG). Nevertheless, the fact that conjugate gradient linear solvers exhibit better load balance between processors in parallel runs with a large number of processors, than the GAMG linear solver, led us to choose the classical PCG solver. We thus favoured good overall parallel performances rather than speed in the case of small numbers of processors, since the goal of RichardsFOAM is to deal with massively parallel computation. However, due to the architecture of OpenFOAM®, it is possible to easily switch to the GAMG linear solver if it is more suitable for a given application (run on only tens of cores for instance). One can refer to Wheeler and Peszyńska [72] for a comparative discussion of preconditioned Krylov methods and multigrid methods in the context of parallel computing.

The choice of the linearization method and of the linear solver were based on the constraints of low memory needs and robustness rather than low computation times, in order to be able to deal with strongly varying hydrological regimes (see the example of computation in a monsoon context in Section 3.2). However, the studies of scaling presented in Section 4 show that the careful implementation of OpenFOAM® leads to relatively low computation times anyway.

Following established procedures (e.g.: [73-75], [47]) we discretized the time evolution with a backward Euler time scheme, because of its well-known stability. For the same reason we use a chord slope approximation for estimation of the capillary capacity (Rathfelder and Abriola, 1994) [76], because this approximation allows a mass conservative use of the pressure formulation of the Richards equation with a classical Picard algorithm. One can refer to [77] for a study of the comparative merits and drawbacks of the chord slope approach. The chord slope approximation is also used in well-known Richards solvers (for instance [17]).

We built a stabilised adaptive time stepping procedure in order to respond to the constraint of stability and of small computation times. There is no need to use small time steps when the variations of hydrologic conditions are smooth, as it would needlessly increase the computation time. However, when sharp changes in hydrologic conditions occur, one needs to immediately diminish the time step to an adequate level. These are the reasons numerous researchers have implemented adaptive time step procedures for the solving of Richards equation (for example, [17], [78], [79], [30], [75], [27], [28], [39]). Here, we have implemented a classical procedure which increases the time step after a given (user specified, 10 is a possible default value) number of time steps with a small number of linearization iterations (3 iterations is a possible default value for the associated threshold), but which immediately decreases the time step when convergence problems are encountered (i.e., when the Picard linearization loop fails, or makes too many iterations; a default value for the associated threshold may be 8 iterations). The decrease/increase is made through the division/multiplication of the current time step with a user specified scaling factor (1.3 is a reasonable default value). In this procedure, a time step with a convergence problem is rerun with progressively decreasing time steps until convergence is reached. This simple approach is the one that Williams and Miller (1999, [78]) call the 'Empirically BAsed Time Step' scheme (EBATS), and it consists *in fine* in selecting an adequate time step serie to achieve a proper functioning of the linearization process. These authors compared the results of this method with those obtained from the multistep BDF (Backward Differentiation Formula) methods for the selection of solution order and step size (Brenan et al., 1996 [80]). According to their work, the use of the empirical method may require longer computation times than the use of multistep BDF methods but leads to similar accuracy. The use of BDF methods also allows prescription of the temporal truncation error, which is not possible when only using the EBATS method. Another way to prescribe the maximum temporal truncation error is the use of a Richardson approach, which evaluates the temporal truncation error through a dual time stepping (e.g.: Belfort et al., 2007 [81], Gasda et al., 2011 [82]). A benchmark of various reactive transport codes, including a discussion of different adaptive time stepping procedures may be found in Carrayrou et al., 2010 [83]. At this point, we must stress that in the case of a Picard linearization method, the use of EBATS time stepping control is strongly recommended in order to ensure convergence of the Picard linearization loop. The additional use of more complex time stepping procedures such as multistep BDF or Richardson methods would ensure a temporal truncation error control, and may consequently reduce the number of Picard iterations. One major drawbacks of the EBATS method is that the control parameters of the adaptive time stepping (e.g.: maximum time step, initial time step, etc) needs to be evaluated empirically, depending on the physical features of the problem considered. The initial and maximal time steps used for the test cases presented in this paper are given as examples of possible parameterizations. One should note that, in the case of a too 'cautious' (small) initial time step, the EBATS procedure will increase its value relatively quickly. Thus, in case of doubt, there is no critical problem with taking a very small (e.g.: 0.1s) initial time step. Nevertheless, the EBATS method has been shown to be quite simple and robust ([78], [83]) which explains its popularity ([27], [28], [30], [39]). , We have also used this simple method, however the additional use of a multistep BDF method or the Richardson method may lead to more efficient solving procedure (truncation error control, optimality of the sequence of time steps; see for example [79], [75], [81], [82]).

Finally, while for each linear system produced by the Picard linearization we have adopted an implicit integration for the pressure terms, we have used an explicit integration for the gravity term (with a simple linear interpolation for evaluating the effective hydraulic conductivity at the cells faces). This has been imposed by the handling of boundary conditions in OpenFOAM®. The explicit formulation of the gravity term was in order to avoid problems encountered while applying the implicit differential operators to the sum of two scalar fields - here pressure and altitude, see equation (1) - because patches for the definition of boundary conditions are defined for both fields. However, the explicit discretization of the gravity term is mitigated by the Picard loop, because at each Picard iteration the pressure field considered for the explicit estimation of the gravity term is that of the previous Picard iteration, and not that of the previous time step. Thus, at convergence of the Picard loop, the expression of the gravity term is almost implicit.

Given the above discussion, we can summarise our implementation of the RichardsFOAM solver in 6 points :

- Picard algorithm for linearization,

- Mixed formulation with chord slope approximation for the capillary capacity,

- Backward time integration scheme for all terms except gravity terms,

- Forward time integration scheme for gravity terms,

- Stabilised adaptive time stepping procedure,

- Diagonal Incomplete Cholesky – Preconditioned Conjugate Gradient (DIC-PCG) for the linear system solver.

One can find more elaborate and recent methods for some of the six points presented above (i.e., Newton linearization instead of Picard linearization, use of multigrid methods, etc), however, we have made our implementation choices in light of our three major constraints (memory use parsimony, maximal stability and robustness, and computation times as small as possible), and this led us to use the simple and classical but reliable and highly scalable methods presented above.

**2.3   Parallel aspects**

The parallel computing in OpenFOAM® is based on the application programme interface MPI (Message Passing Interface) procedures, with the use of the mesh partitioning approach. It is based on a Single Program – Multiple Data (SPMD) approach: each core makes the computations on a geometrical fraction of the resolution domain, and runs the same programme on this sub-domain that would be used on the whole domain in serial computation. The boundary condition at the edge of each sub-domain is the physical boundary condition if the edge is also an edge of the global domain, or an inter-cores communication boundary condition if the edge is interior to the global domain.  The mesh partitioning approach is the most frequently used parallelisation strategy in unsaturated flow modelling (e.g.: [35-38], [43], [46], [47]), although other approaches exist, such as parallelization by loop decomposition (for example, [34], [39], [40]). Nevertheless the loop decomposition approach is used only for a small number of cores (one tenth or less).

In order to achieve high parallel performances one needs to ensure that two main requirements are satisfied: a good computational load balance between cores, and a sufficiently high ratio between computational operations and inter-cores communication operations.

The computational load must be fairly balanced between each computing process (core), otherwise the slower core will delay the whole computation. The choice of the mesh partitioning method is critical in this regard. In the framework of OpenFOAM®, different partition methods may be used: simple, hierarchical, scotch, metis, or manual (Cf. [84]). In this work, we used the simple method because the simplicity of the geometry of the cases presented in this paper allows it to reach perfect load balance in terms of the number of mesh cells by process (core). The choice of the linear solver is also important: as discussed above, some very efficient linear solvers like GAMG will cause problems of load imbalance in massively parallel runs. For this reason we preferred a PCG linear solver.

The ratio between computation and communication needs to be high enough, otherwise the computations within a given core will be ended before they can be sent and received from one process (core) to another, hence periods of inactivity (wait) will be of the same magnitude as the period of computing. There are two ways of limiting communications weight: (i) limiting the size of interfaces between sub-domains of the mesh (control the amount of communication) and (ii) keeping a high ratio between the number of elements of a sub-domain and the number of elements at the interfaces (more computation than data transfer). The first point must be achieved by a careful choice of the mesh partitioning method. The second point requires a large enough cell number by process (core) : small scale computations tend not to be made with good parallel performances (e.g., ref. 35). A final point to be considered in the case of a non-linear equation such as the Richards equation is the method of implementing the exit test of the linearization loop. The criterion for exit of this loop must be evaluated globally (i.e., for the whole domain, including all subdomains). The global evaluation of the criterion of exit implies an additional global reduction operation at each iteration of the linearization loop, but if the criterion is estimated locally (only on the sub-domain associated with a given core for the computation done by this core), the convergence of the computation will be mesh partitioning – dependent.

**2.4   Hardware and system :**

For this work, all parallel computations were run on the Cluster "Hyperion" of the Scientific Computing Centre of University of Toulouse (CALMIP, www.calmip.cict.fr). The distributed memory part of Hyperion is a cluster Altix ICE 8200 SGI® purchased in 2009 (reach #223 TOP 500 in November 2009). The cluster gathers 368 computing nodes (2944 cores), each node includes 2 quad-core NEHALEM Intel® processors 2.8Ghz and 36GB of RAM. The fast interconnection of the cluster involves DDR Infiniband Technology, via two physical Infiniband interconnects and so two IB links are associated with each computing node. The theoretical bandwidth at node level is thus 40Gb/s instead of 20Gb/s for a single IB DDR link. Moreover at the application level, Message Passing Interface communications use fast interconnection fabric, although one of The IB DDR links (and always the same one) is devoted to I/O traffic (read/write data on file system). This is important to properly analyse subsequent performance. Finally the topology of these two distinct interconnections is a hypercube topology. We used the local Lustre file system to bench I/O performances of RichardsFOAM. The characteristics are the following: 200TB of disk space, 4 Object Storage Servers, one Meta-Data Server and a measured bandwidth performance of 3GB/s. Lustre is a well-known and widespread parallel file system for high performance computing, thanks, among other things, to its ability to handle a large number of clients (or nodes). It is used, for instance, in the system CURIE thin nodes at CEA (France) (#11 in November 2012's TOP500 ; CURIE has more than 4,000 thin nodes). In addition to technical aspects related to distributed computations, we also used a fat node with a large amount of shared memory (3 TB) to handle some steps related to pre-processing, for instance, to construct and partition meshes in some huge cases.

# 3 Applications

In this section we will present several applications of RichardsFOAM in order to point out the capabilities of this solver. First, a code-to-code validation is presented for a transient case. A field data set obtained from an agricultural area in a monsoon affected region is then studied, in order to illustrate the ability of RichardsFOAM to handle field data chronicles and to deal with strongly contrasting hydrological regimes. In addition, in Appendix A (electronic supplementary material) one can find an analytical validation in a very simple steady state case. We do not present further validation cases because as mentioned in the implementation part (section 2.2) we use well known, simple and classical (but reliable) numerical methods for the solving of Richards equation.

## 3.1 Code-to-code validation

In order to proceed to a validation of RichardsFOAM in transient conditions, we must perform a code-to-code benchmark on a simple 1D infiltration case with the well-known Richards equation solver Hydrus-1D (e.g.: [27], [28]). The mesh is a regular grid of 1cm mesh cells for both codes, the initial time step is 5 minutes and the maximum time step is 1 hour. We consider the infiltration of water in a loam column ($K_s = 2.89 \times 10^{-6}$ m.s$^{-1}$ ; $\alpha = 3.6$ m$^{-1}$ ; $n = 1.56$ ; $\theta_s = 0.43$ ; $\theta_r = 0.078$) of 1 metre thickness. The initial condition is a constant pressure field of -1 m. The bottom boundary condition is a free-drainage condition and the top boundary condition is a fixed pressure equal to 0.01m. One can see in Figure 1 the features of the problem considered.

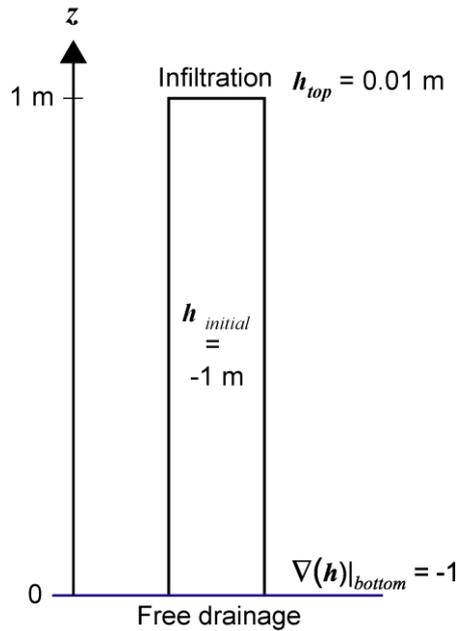

Figure 1: Geometry, initial conditions and boundary conditions of the problem considered for code-to-code validation.

We have computed the time and space evolution of the pressure field within the soil column until a steady-state is reached with both models, Hydrus-1D and RichardsFOAM. One can see a comparison between the results of each models in Figure 2, in terms of average water pressure in the soil column (i.e., integral over the column of the pressure field divided by the volume of the column) and in terms of top water flux as a function of time.

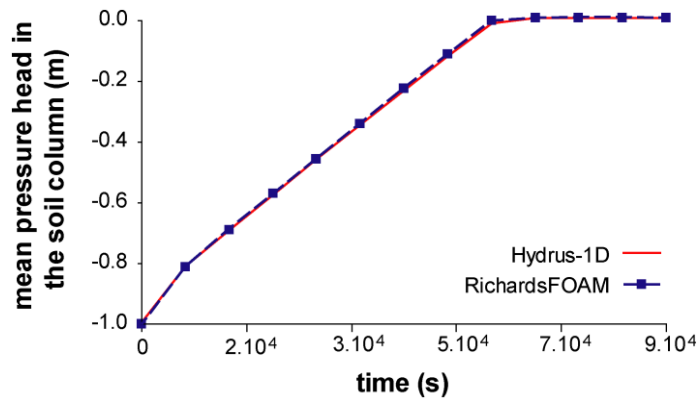

Figure 2: Comparison of Hydrus-1D results and RichardsFOAM results for a 1D transient case.

There is good agreement between both computations. For example, the absolute values of the relative differences between the computed fluxes are strictly lower than 1.5%.

**3.2 Field test case: multilayer soil column with time variable boundary condition**

In this sub-section we want to illustrate the ability of RichardsFOAM to handle time variable input data provided by the observation of natural systems, as well as to run simulations of flows in contexts with abrupt changes in hydrological regimes. Indeed, strongly varying hydrological regimes lead to rough numerical problems, due to the steep pressure fronts which occur in such conditions (see for example [85]). We consider here a data set acquired in South India, from a turmeric field in the state of Karnataka [86]. In this region, the monsoon climate causes highly contrasting dry and wet seasons. Rain, irrigation and potential evapotranspiration were measured on a daily basis during 2011, as was the water content in the soil at 40cm depth and 120cm depth. Actual evapotranspiration was estimated using a Thornthwaite balance with the assumption of an easily usable water storage of 100mm. This allows the use a time varying Neumann top boundary condition which takes into account combined precipitation and actual evapotranspiration (using the groovyBC tool [87]). This approach requires pre-processing of the data before the implementation of the top boundary condition, and allows only a rough estimate of this boundary condition A more integrated way of dealing with these top boundary conditions would lead to consideration of additional mass balance equations for estimating actual evapotranspiration on the basis of potential evapotranspiration, by considering the hydric state of the first soil layers 'on the fly', but this is beyond the scope of this work. An Auger hole investigation established a rough representation of the soil column, with the identification of three layers in the first 2 m. The bedrock is at an average depth of 10m in this area. One can see in Figure 3 the configuration of the considered soil column.

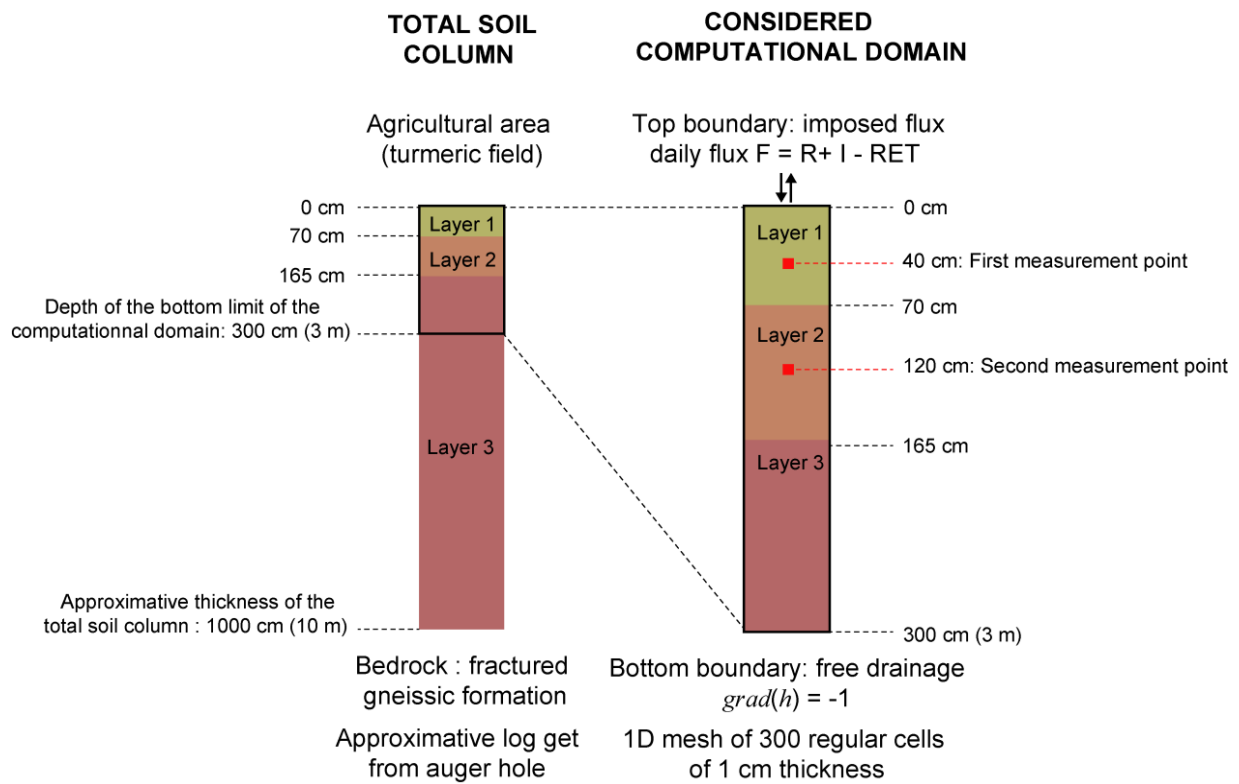

Figure 3: Soil column considered for the real dataset fitting exercise, with the associated computational domain.

The rain, irrigation and actual evapotranspiration data were used as input for RichardsFOAM, and a fitting exercise was undertaken to get an acceptable correspondence between computed and measured water contents. The mesh used here was a 1D-mesh with 1cm thick cells. The precision for the PCG linear solver was set to $10^{-11}$m, and to $10^{-9}$m for the Picard linearization. The initial time step was set to 1 minute and the maximum time step was set to 1 hour. It is important to note that, in order to get an efficient and accurate modelling, these numerical features need to be tuned for each application under consideration, depending on the precision required for the results, the spatial and temporal scales under consideration, and so on. We limited our computational domain to the first three metres of the soil column since sensitivity tests have shown that the values for the water content of the first two metres of soils (where the measurements take place) have a very weak dependency on the flow properties beyond 3 metres depth (data not shown). We also make the assumption that run off could be neglected on the considered plot. Therefore, this exercise is only an illustrative example and does not pretend to deal with the full complexity of water flows in the considered area. Nevertheless, the measured dataset and the comparison between numerical results and observations are presented in Figure 4.

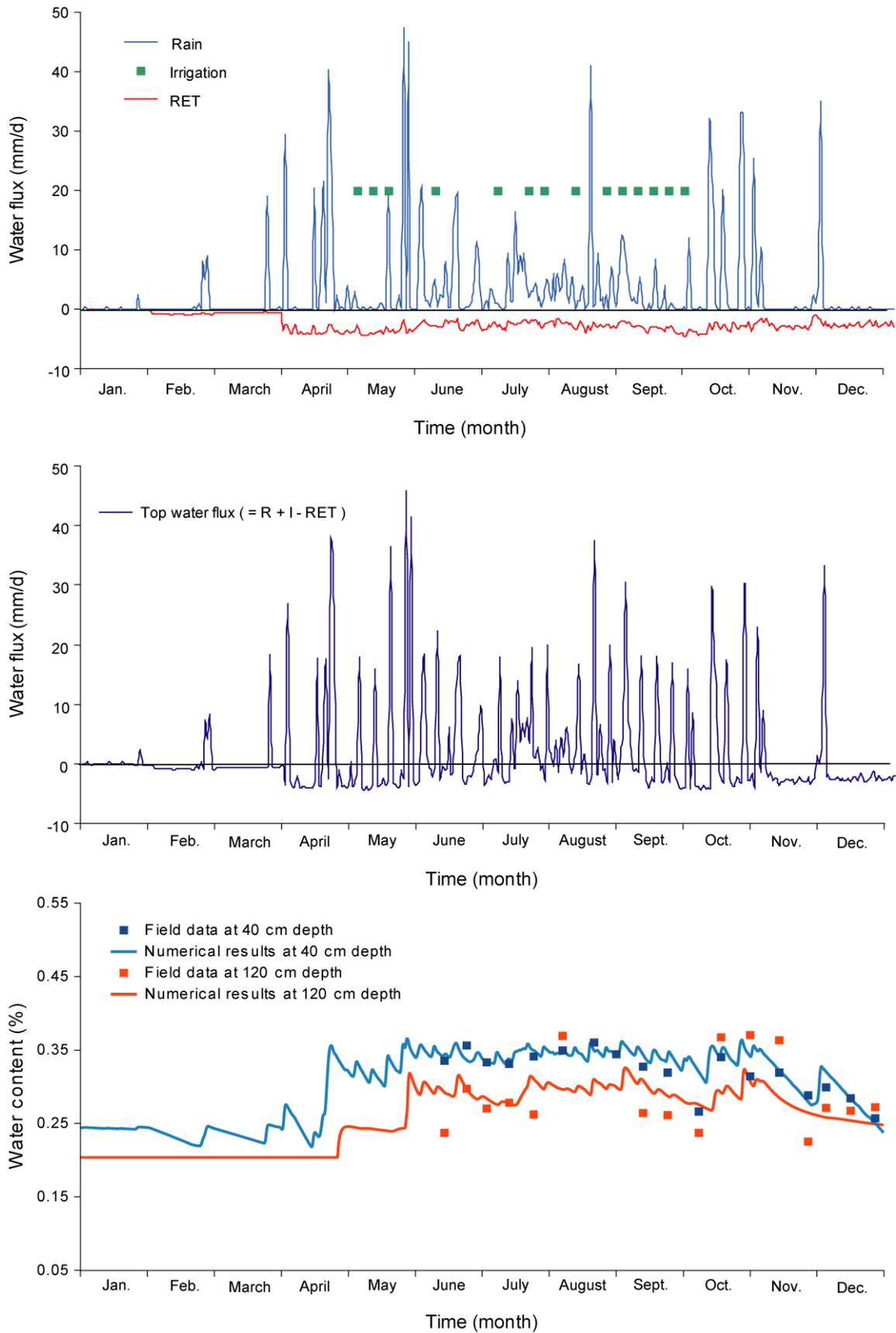

Figure 4: Data set used for the fitting exercise and comparison of numerical results and observations.

One can see in Figure 4 that we have a strongly varying hydrological regime, with a dry season from November to March and a wet season from April to October, which is typical of an area with a monsoon climate.

The fitting of the parameters has been done on the soil hydrodynamic properties of the van Genuchten soil model. The set of fitted parameters is presented below:

$$K_s^{\text{layer 1}} = 10^{-3}\,\text{m.s}^{-1}\ ;\ \alpha^{\text{layer 1}} = 2.3\,\text{m}^{-1}\ ;\ n^{\text{layer 1}} = 1.2\ ;\ \theta_s^{\text{layer 1}} = 0.45\ ;\ \theta_r^{\text{layer 1}} = 0.08$$
$$K_s^{\text{layer 2}} = 10^{-5}\,\text{m.s}^{-1}\ ;\ \alpha^{\text{layer 2}} = 2.1\,\text{m}^{-1}\ ;\ n^{\text{layer 2}} = 2.4\ ;\ \theta_s^{\text{layer 2}} = 0.5\ ;\ \theta_r^{\text{layer 2}} = 0.2 \quad (5)$$
$$K_s^{\text{layer 3}} = 10^{-6}\,\text{m.s}^{-1}\ ;\ \alpha^{\text{layer 3}} = 1.7\,\text{m}^{-1}\ ;\ n^{\text{layer 3}} = 1.9\ ;\ \theta_s^{\text{layer 3}} = 0.55\ ;\ \theta_r^{\text{layer 3}} = 0.15$$

We obtain reasonably good agreement between numerical results and observations, with an averaged relative difference of less than 2% (and an average of the absolute values of the relative differences less than 8%), which is lower than the precision of the measurements themselves (about 10-20%). One can see that the first soil layer has quite high hydraulic conductivity. This could be due to the fact that this fitted hydraulic conductivity takes into account the existence of macropores in this first layer.

Overall, RichardsFOAM shows good capacity to handle real climatic data set in a layered soil column, in strongly varying hydrological regimes.

## 4  Study of the parallel performances

All performance results are based on computing experiments that were run on the system described in Section 2.4. As previously mentioned we used NEHALEM Intel® 2.8Ghz cores that deliver more than 11Gflops/s. Peak performance for 1024 cores, which is the highest bound of this parallel study, is thus above 11 264Gflop/s (11.26Tflop/s peak). As our interest is in the parallel performances we can achieve (see subsequent section with strong and weak scaling), we will also focus on the I/O impact on parallel performances. Indeed I/O operations (e.g.: writing of results) may be a critical source of non-scalability, but parallel numerical tools like RichardFOAM must of course provide results to be analysed in order to be useful, and thus I/O are necessary.

### 4.1 Conditions of runs

Most of the runs undertaken on the cluster 'Hyperion' (cf. section 2.4) have been performed in production conditions, with many other jobs running on the cluster at the same time. A job scheduler handles the scheduling of the workload on the system. A computing resources request is expressed in terms of nodes (cores+memory) and elapsed time. Resources that are finally allocated are exclusives. Nevertheless, there are no constraints, and so no guarantees, of any interconnection topology affinity in the group of computing nodes that are selected by the job scheduler. Eventually, even if computing resources are allocated in an exclusive mode, this will not be the case for the Lustre File system (cf. Section 2.4). More precisely, its bandwidth is potentially shared by all applications (jobs) currently running on the system. In the subsequent sections, we will implicitly use the mapping of one process MPI for one physical core. For each run, we complete all cores of a socket (quad-core), and each socket of the computing node (bi-socket nodes).

### 4.2 Strong scaling

In this part we assess the ability of RichardsFOAM to speed up computation related to parallel computing. The method adopted is a strong scaling study: we solve the same problem on more and more cores and we study the evolution of the associated speed up. The case considered is an infiltration of rainwater over 10 days on a 3D homogeneous slope of dry loam (initial water pressure: -10m, soil properties identical to those considered in Section 3.1), with a river at the bottom. We consider a 10m thick slope, with a length and width of 1.7 km. The total associated surface is 2.9km$^2$ and the total volume of the considered domain is about 29 millions m$^3$. The slope is inclined at 20° in a direction orthogonal to the river, and the river is inclined at 3°. One can see the considered geometry in Figure 5, the boundary conditions and initial conditions.

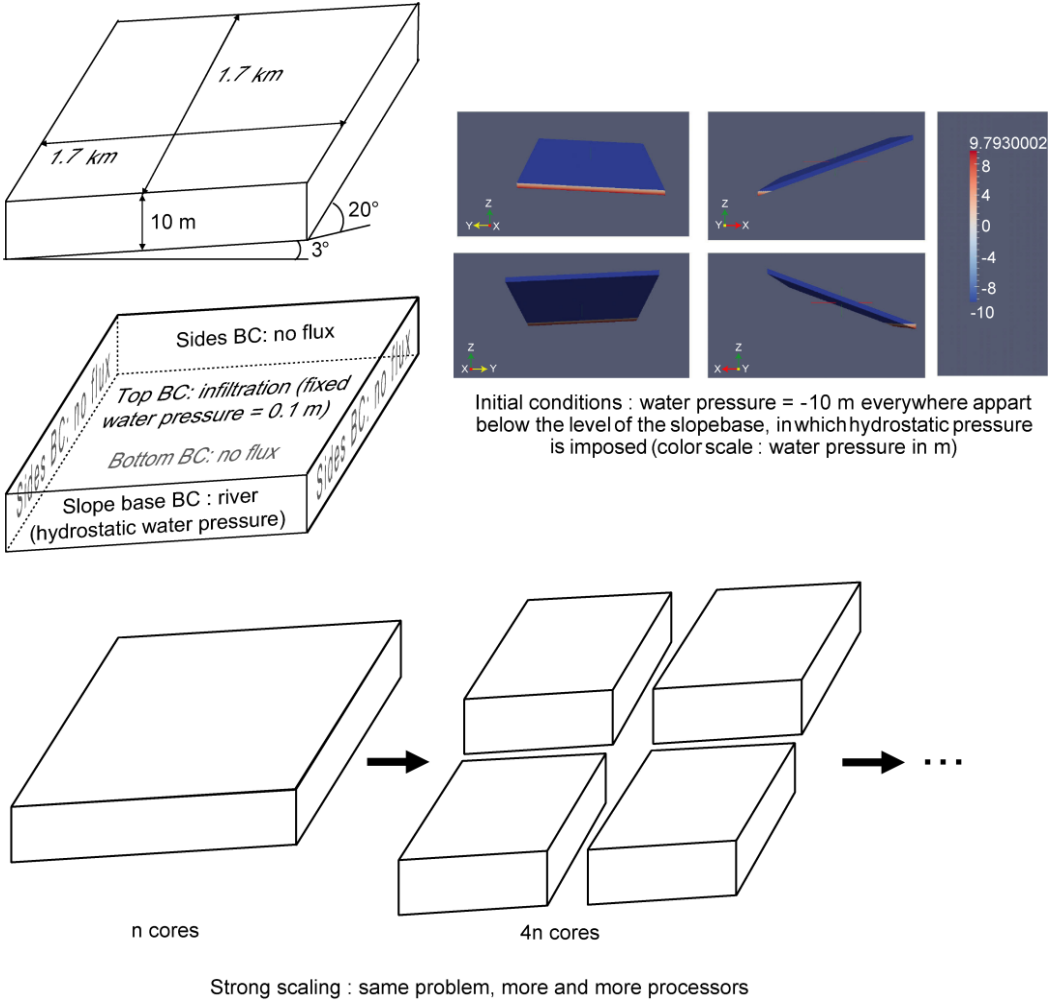

Figure 5: Domain, boundary conditions and initial conditions for the strong scaling exercise (drawings not to scale).

We have chosen to consider a case with an homogeneous soil because a random and spatially uncorrelated heterogeneity of the medium does not impact the scalability when using a Picard-PCG approach of solving (see an illustration of this point in electronic supplementary material, Appendix B). In case of a medium heterogeneity with a spatial correlation (e.g.: layering, lenses of material with different mean hydraulic conductivity, etc) at the length scale of the size of the sub-domains, or in case of occurrence of steep fronts of water pressure within the medium, some computation sub-domains may experience strong penalizing numerical conditions compared to the other sub-domains. Then the scalability may be damaged We think however that it would be related to problems of mesh partitioning optimality, not to the solver implementation itself (see [70]). Moreover, the implementation of RichadsFOAM limits the impact of the steepness of the considered problem on the scalability (see an illustration of this point in electronic supplementary material, Appendix B). Thus we have also considered smooth hydrological regimes (in particular, no transition between saturated and unsaturated conditions during the infiltration) for this strong scaling analysis as well as for the following weak scaling analysis.

A careful convergence study (criterion: less than 5% of maximum relative difference compared to the free mesh dependency and free precision dependency results, data not shown) leads to the conclusion that acceptable results in term of average soil water content in the slope can be reached with a regular mesh with cells of 2m × 2m in the horizontal plane and 0.2m in a vertical direction (which leads to a mesh size of about 36 million cells), a precision of 0.1mm for the PCG solver and a precision of 1 mm for the Picard loop. The initial time step is 0.1s and the maximum time step is 30 minutes. Indeed, this prior step in the convergence study is necessary in order to obtain accurate quantitative results (e.g.: [5]). With such features, it is possible to catch the transient infiltration within the slope. The main difficulties here come from the contact between the saturated zone (the river) and the unsaturated slope and from the high moisture contrast on the infiltration front. For the building of the mesh we have used a fat node of 3TB of shared memory. The conditioning of the case (initial and boundary conditions) has been performed using the "swak4foam" library.

A strong scaling exercise using from 16 cores (2 nodes) to 1024 cores (128 nodes) was performed for this case, and the results in terms of speed up and parallel efficiency are shown in Figure 6.

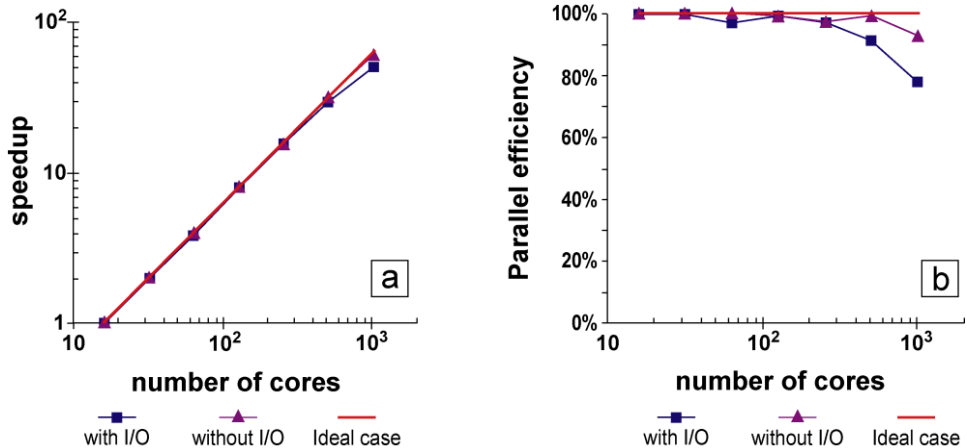

Figure 6: Speed up (a) and parallel efficiency (b) with and without I/O for the strong scaling exercise (computation times with 16 processors: about 1h15mn; with 1024 processors: between 1mm and 2mn - mesh size: about 36 million cells).

The elapsed computation time was about 1h15mn with 16 cores (number of mesh cells per core: about 2.3 million) and between 1 and 2mn with 1024 cores (number of mesh cells per core: about 35 000). In order to assess the impact of I/O on parallel performances of the solver, two strong scaling curves have been created : one with written results for each day of rain, the 'with I/O case', and one with written results only at the end of the 10 days of rain, the 'without I/O case'. Note that the amount of data to be written in one single time step with I/O is roughly 2.5 GB. In the 'with I/O case', 10 computing time steps involved such I/O.

To compute the speed up and parallel efficiency we use the total elapsed computing time for each run. One can see that the speed up is very good at up to 1024 cores. It is worth noting that the I/O operations have an important impact on parallel performances only for the highest number of cores: we have more than 90% parallel efficiency with 1024 cores without I/O, while we have just less than 80% of efficiency with I/O. In Section 4.4 we will focus on I/O analysis in order to identify some of the reasons for the loss in parallel efficiency. Thus, in the case of massive parallel computations with RichardsFOAM, the frequency of writing the current solution has to be carefully chosen to avoid a collapse in the parallel efficiency.

## 4.3 Weak scaling

In order to evaluate the ability of RichardsFOAM to deal with large scale problems, we performed a weak scaling study on cases with similar features to those used for the strong scaling exercise, apart from the soil depth (6m for the weak scaling exercise) and the boundary condition at the bottom of the slope (here zero pressure). In this weak scaling study we considered a sequence of problems with increasing sizes in terms of number of mesh cells, and each element of this sequence was solved with an increasingly important number of cores so that the number of mesh cells per core stayed constant along the sequence (number of mesh cells per core: about 130 000). Figure 7 shows the adopted method of building the different cases associated with each element of the sequence.

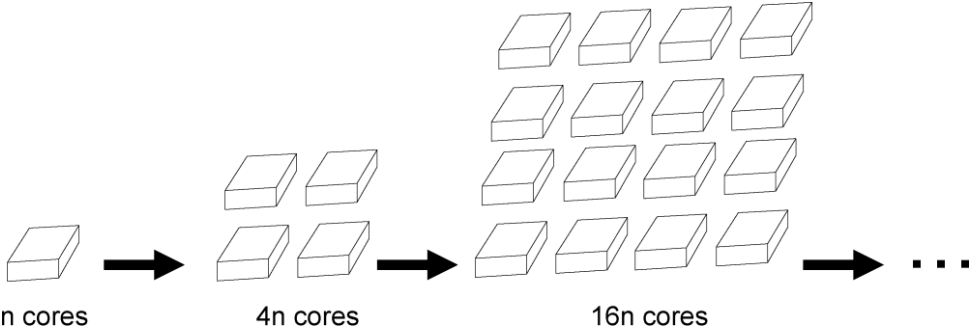

Weak scaling : bigger and bigger problems as well as more and more processors

Figure 7: Considered cases for the weak scaling exercise.

From a parallel analysis point of view, this is a way to see whether the ratio between computation and communication is kept steady.

For this weak scaling exercise we considered two cases as well as for the strong scaling exercise: a case with I/O, in which results were written for all days of physical time, and a case without I/O, in which results were written only at the end of the computation. One should note that in this weak scaling exercise the whole pre-processing procedure (mesh building, problem conditioning, initial conditions setting) was done in parallel, without using a fat node with a large amount of shared memory.

One can see the results of the weak scaling exercise in terms of parallel efficiency in Figure 8.

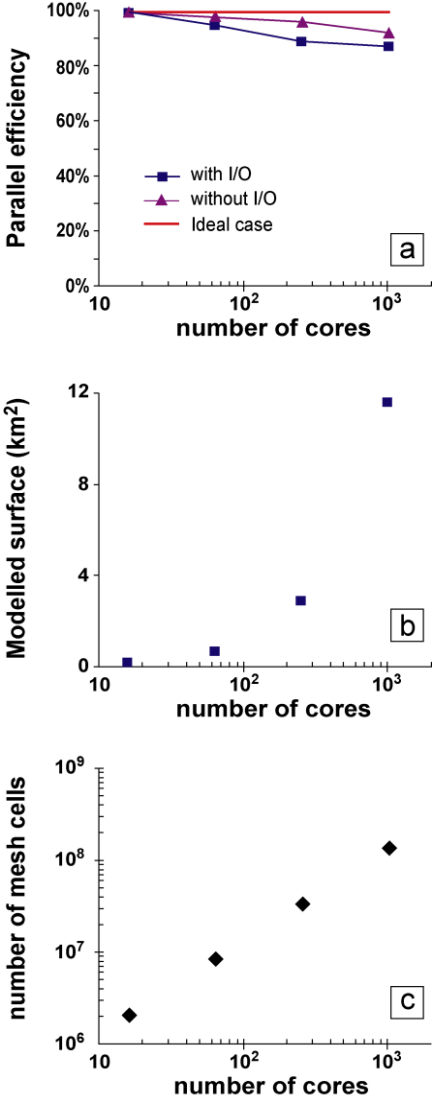

Figure 8: Weak scaling study results and features: parallel efficiencies (a), modelled surfaces (b) and mesh sizes (c). Average computation times: 5mn for a physical time of 10 days.

To compute the speed up and parallel efficiency, we again consider the total elapsed computing time for each run. In a weak scaling approach the speed up is ideally constant and equal to one. We tried to determine whether time to solution was kept steady as we increased the size of the problem and accordingly the number of cores. A deviation from ideality lead to a speed up lower than one, thus the speed up and the parallel efficiency are equal. This is why we show only the parallel efficiency curves here. In order to illustrate the ability of RichardsFOAM to deal with large scale problems, we also show in Figure 8 the physical surface of the slope for each problem, and the associated mesh size. The computation times for this weak scaling exercise were about 5mn (for a physical time of 10 days, as in the strong scaling study). Here, unlike the strong scaling case, the amount of data to be written increased with the number of cores involved (in fact with the size of the mesh used for a given number of cores). The amount of data to be written in one single time step in the 'with I/O' case was 0.7GB for 64 cores (8 nodes), 2.8GB for 256 cores (32 nodes) and 11.2GB for 1024 cores (128 nodes).

The performances were satisfactory in terms of parallel efficiency, which remained over 90% even for the greatest number of cores. We experienced the same trends as in strong scaling with an I/O impact on parallel efficiency, that is, a drop to 80% in the largest case with 1024 processors. Taken together, the strong scaling study and the weak scaling study show that RichardsFOAM may allow a simulation to run on large space scales (km$^2$) and time scales (decades to century) with acceptable computation times.

### 4.4 Impact of data writing

In this section we analyze the reason of parallel efficiency drop at high number of cores. In Figure 9 we plot the computation times at iteration level for the strong scaling exercise in the 'with I/O' case for three different numbers of cores: 64, 256 and 1024 cores.

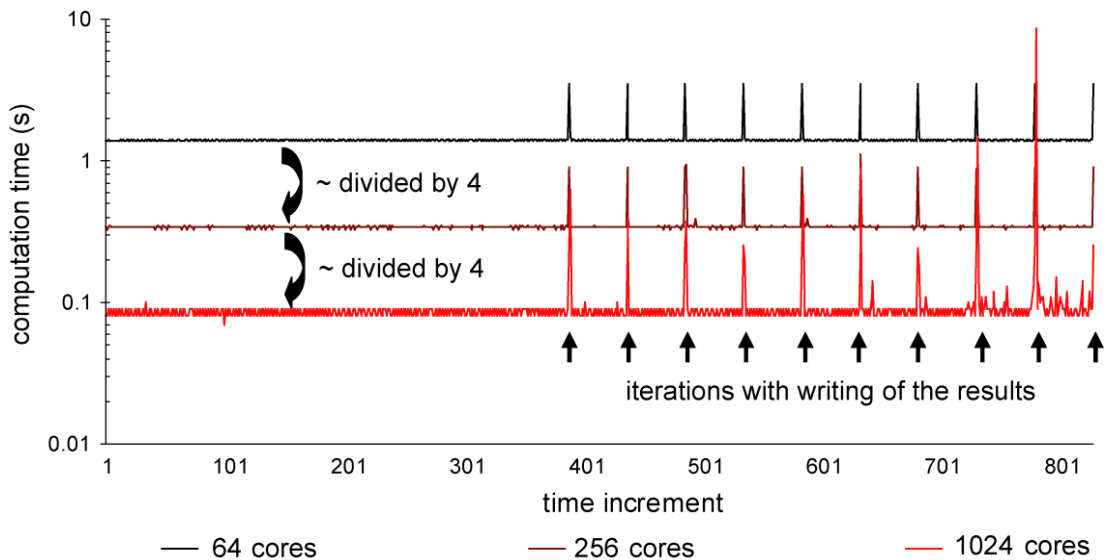

Figure 9: Computation times by time increment for the strong scaling exercise with I/O.

It can be seen on this graph that, in general, at the iteration level the computation time is divided by approximately 4 when the numbers of used cores are multiplied by 4. Small oscillations occur in the computation times for a given number of used cores. These small oscillations may be related to the production context on the cluster (cf. Section 4.1). Communications between the cores are slightly influenced by the surrounding flow of data associated with the other jobs simultaneously running on the cluster. This is most certainly true with large oscillations at the end of the 1024 cores case.

It can also be seen that, for each time step corresponding to the I/O phase, there is a significant increase in the computation time, especially for the high numbers of cores. Since these increases in computation time are far more important than the oscillations observed between the iterations with no writing phase, we strongly suspect that the global drops of efficiency observed for a great number of cores in the scaling studies (see Sections 4.2 and 4.3) are mainly related to I/O effects. The question here was to determine whether this lack of performance was related more to application algorithm than to the performance of the Lustre

File system (see Section 2.4). From the application algorithm point of view, the important features for the subsequent analysis are as follows. At a time step enduring the writing of data (I/O), after convergence, each core is writing its own data fields corresponding to the unknowns of the underlying sub-mesh. Each core is writing its own data in a specific file, thus each core is writing independently and the amount of data is balanced by the balanced partitioning of the mesh. This is an efficient way, at least from the algorithmic point of view, to perform the global I/O in parallel and for a reasonable amount of cores (nodes). One should note that the post-processing of this data may require the reconstruction of one single object from several data files.

We thus had to turn to file system consideration in order to try to understand the drop in parallel efficiency with I/O. One must bear in mind that we are dealing with a loaded global computing system (for the period in which we ran simulations the average global load of the cluster was above 80%). Hence the bandwidth capacity of the file system cannot be fully available for our runs. It is shared between all jobs currently running on the system and potentially reading or writing.

A power outage gave us the opportunity to make some runs on an empty system, which means that the whole system (computing resources and file system) was devoted to the runs. We therefore reran some of the scaling tests shown in Sub-sections 4.2 and 4.3 in a computing system configuration where we were sure that the entire capacity (i.e., bandwidth) of the file system was available for our runs. Results are illustrated in Figure 10.

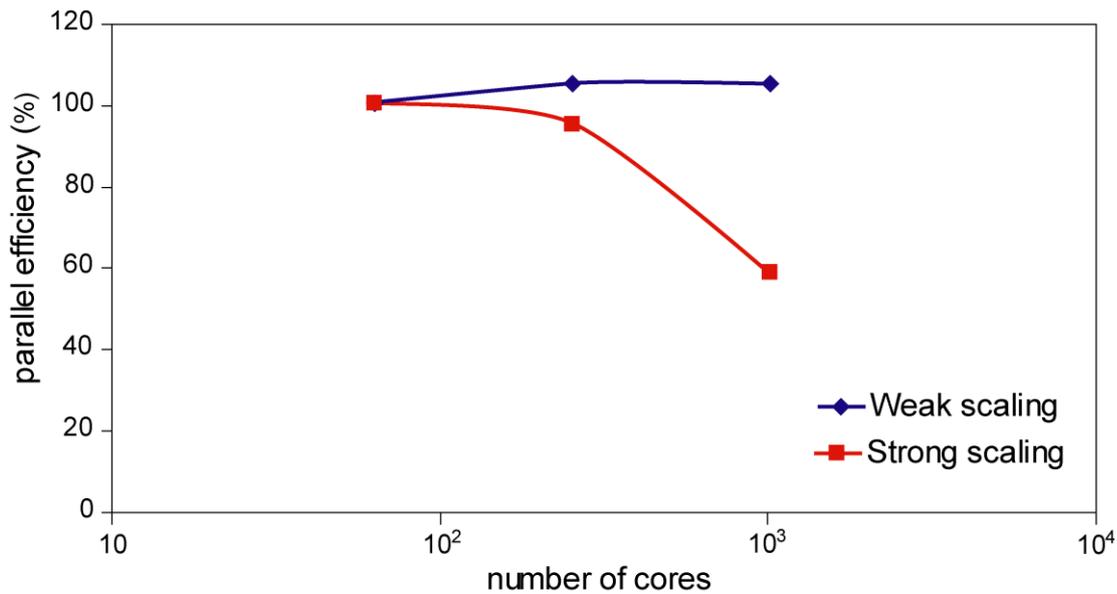

Figure 10: Parallel efficiency of I/O in the last iteration of the computations done with a dedicated Lustre file system.

In this graph, we focused on the last iteration with I/O in both the strong and weak cases, and more precisely on the overhead due to writing data. We then calculated parallel efficiency classically, by considering the run with 64 cores as the reference run, both for the strong scaling and the weak scaling. It can be seen that we had two different parallel behaviours corresponding to strong and weak scaling cases. In the weak scaling case the parallel efficiency was super-linear around 100%, whereas in the strong scaling case, parallel efficiency dropped to 60% at 1024 cores. As these performance results are the best we can achieve on our computing system, we can conclude that the lack of parallel performance pointed out in Figure 9 is due to production conditions during the runs (a highly loaded computing system), because with a devoted computing system we observed a good scalability for the iterations with I/O operations, and not massive drops as with a computing system in production. The different behaviours observed between the strong scaling case and the weak scaling case can be explained by the design of the Lustre file system. For the drop to 60% in strong scaling, the observed behaviour can be related to the ratio between the meta-data and the data for a given file. In the weak scaling case this ratio is constant, whereas in the strong case it decreases constantly. Overhead due to managing meta-data can thus be of the same magnitude as writing data itself, which explains the lack of parallel efficiency. On the other hand, for the slightly super-linear behaviour, it can be seen that in addition to any bufferisation/cache effects, an higher number of nodes (i.e., Lustre clients) may be allowed to catch the full bandwidth of the Lustre file system, if the global amount of data is above a threshold related to the tested Lustre file system (cf. Section 2.4).

I/O issues play a significant role in the parallel performance of applications in general, and more particularly for RichardsFOAM. However, we showed that in some cases (weak scaling), the underlying capacity (bandwidth and parallelism) of the file system may erase the negative impact of I/O on performance. In these cases, the drop in parallel efficiency (illustrated in Section 4.3) is solely due to the production context.

## 5   Conclusions and perspectives

In this work we developed a massively parallel open source solver for Richards equation, the RichardsFOAM solver. The advantage of building such a solver within a generalist open source CFD platform such as OpenFOAM® is that it allows easy collaborative testing and improvement by large groups of developers, and can benefit from the various developments in numerical tools developed by the OpenFOAM® community, which embraces a wide range of aspects of computational continuum mechanics. Validation cases have been proposed and application to a real data set acquired in a region with a monsoon climate illustrated the ability of RichardsFOAM to deal with strongly varying hydrological conditions. The solver has demonstrated state of the art parallel performance, both for strong and weak scaling. The conditions for obtaining such performances have been discussed, and one of the important points identified is the good handling of I/O operations, consisting in i) minimal writing and

ii) using an adequate file system. This I/O-related loss of scalability has been shown as mainly due to the production context of the cluster used rather than to the implementation of RichardsFOAM itself or of the OpenFOAM® tool box.

Although the field of applications for a massively parallel solver for Richards equation is broad (environmental engineering, water resources management, etc), RichardsFOAM was primarily developed in order to provide a quantitative and predictive tool to give relevant hydrological input data to continental weathering modeling (for instance, [1], [88]). The parallel performances of RichardsFOAM will allow, in the near future, application to decades to century time scale modelling of water fluxes in the critical zone of experimental watersheds with surfaces of a few square kilometres, in order to study the impact of global changes on weathering processes. Many experimental watersheds in this range of scales are currently monitored in order to study weathering processes: for instance in a tropical climatic context such as the Mule Hole watershed (South India, 4.3 km$^2$ of surface [89]), Maddur watershed (South India, 7.2 km$^2$ of surface [90]) or Nsimi watershed (Cameroun, 1.1 km$^2$ of surface [91]). In boreal climatic contexts, catchments at the corresponding spatial scale are also monitored in Central Siberia (Kulingdakan watershed, 12 km$^2$ of surface; e.g.: [92]) and thus should also be the subject of studies with the developed tool, for example to characterise water processes during the summer season. Such applications of RichardsFOAM on experimental watersheds will be the scope of future works devoted to the modelling of water fluxes on continental surfaces. Further tests of the computational performance of the open source parallel RichardsFOAM solver may deal with new test problems with highly non-linear and highly spatially correlated heterogeneous unsaturated flow problems in one and several dimensions. This will allow testing the overall performance of the numerical solver (combining assessments of parallelization efficiency, Picard iterations, number of time steps, etc).

Finally, the good parallel performances obtained with RichardsFOAM illustrate the benefit of developing a more complete tool for modelling transfers (of surface waters as well as of subsurface waters, of solutes, of energy) in continental surfaces in the framework of OpenFOAM®. This will be the subject of future works devoted to the set up of mechanistic modelling of hydrogeochemical processes at the watershed scale.

**Acknowledgments**

We are grateful to two anonymous reviewers for their helpful and constructive comments. This work was supported by a Paul Sabatier University starting grant. Partial support from BIO-GEO-CLIM grant No 14.B25.31.0001 is also acknowledged.

Finally we would like to thank the whole team of the CALMIP cluster.

**Appendix A**

In this appendix we show an analytical validation of RichardsFOAM. We consider here a very simple 1D configuration inspired from the analytical solution of Richards equation presented in [35]. We consider a homogeneous soil column of 1 meter in height, with a water table at the bottom ($h = 0$m) and various imposed flux densities at the top (noted $q$, [m.s$^{-1}$]), either in infiltration and evaporation conditions. One can see in Figure A1 the considered configuration.

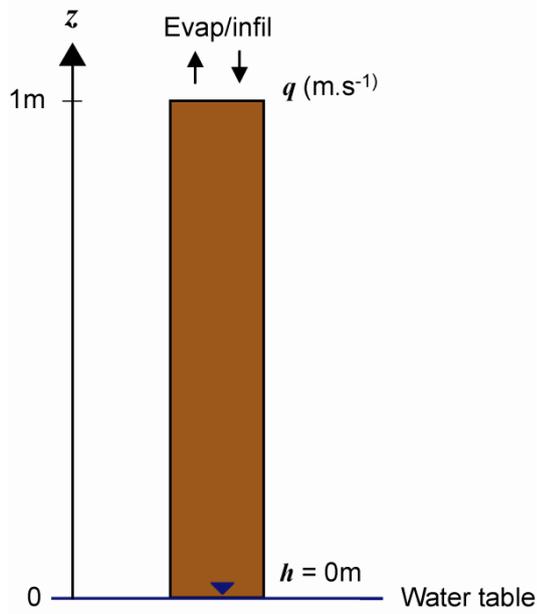

**Figure A1: Considered 1D configuration for the 1D analytical validation.**

We consider a steady state problem. In order to be able to obtain an analytical solution, we use Gardner's correlation [93] for the relative hydraulic conductivity (Eq. A1).

$$K(h) = K_s e^{\alpha h} \tag{A6}$$

with a parameter $\alpha$ [m$^{-1}$] and a saturation hydraulic conductivity $K_s$ [m.s$^{-1}$] which depend on the soil properties. In 1D and steady state conditions, one can rewrite Richards equation as follows:

$$\frac{d}{dz}\left(K(h)\frac{dh}{dz}\right) + \frac{d(K(h))}{dz} = 0 \tag{A7}$$

In the case of Gardner relative hydraulic conductivity, the use of the Kirchhoff transform allows the following analytical solution for the considered case to be established (valid for $q < K_s e^{-\alpha}/(e^{-\alpha}-1)$):

$$h(z) = \frac{1}{\alpha} \ln\left[\frac{1}{K_s}\left(q + (-q + K_s)e^{-\alpha z}\right)\right] \tag{A8}$$

Here we have arbitrarily considered a saturated hydraulic conductivity $K_s$ of $10^{-6}$m.s$^{-1}$, and a parameter $\alpha$ equal to 0.06m$^{-1}$. One can see in Figure A2 the comparison between the analytical solutions and the numerical results obtained with RichardsFOAM, with a mesh of 100 regular grids of 1cm width, a precision of $10^{-10}$m for the PCG solver and a precision of $10^{-8}$m for the Picard loop. The initial time step was 1 second and the maximum time step was 3600 seconds.

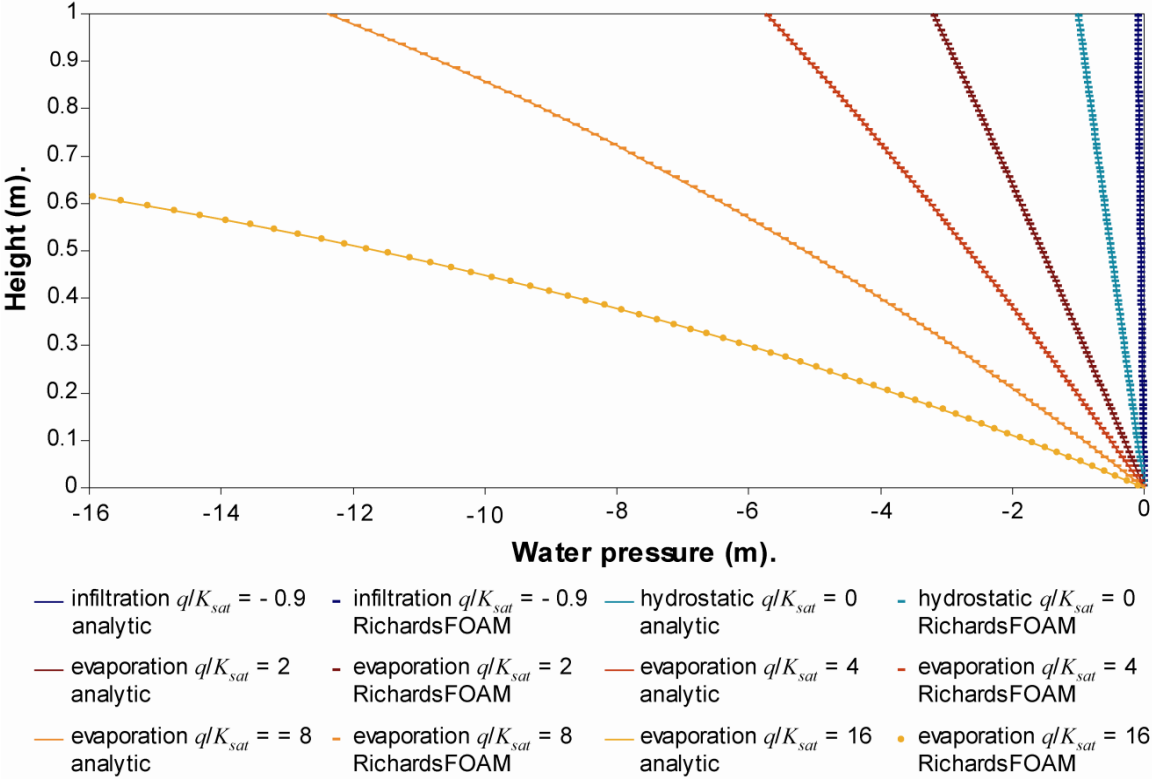

**Figure A2: Comparison between analytical solutions (continuous lines) and numerical results (symbols) of RichardsFOAM for a 1D steady-state case.**

The comparison has been made for a range of fluxes which were compatible with no occurrence of saturated areas in the soil column (Gardner's equation would lead to non-physical behaviour for a positive pressure) and with a flux just below the limit of validity of Eq. A3. A positive flux density is associated with evaporation (upward flux) and a negative flux density is associated with infiltration (downward flux). The case of a zero flux is the case of hydrostatic equilibrium. There is good agreement between RichardsFOAM results and the analytical solutions, which can be considered a satisfying validation.

**Appendix B**

The cases considered in the presented scaling studies involve homogeneous soils and smooth infiltration regimes (no saturated/unsaturated transition). The question arises to which extend the scalability observed in these simple cases can be damaged in the case of heterogeneous soils and steep hydrological regimes ? Batzoglou et al., 1992 [94] showed that the Picard-PCG approach adequately solved Richards equation when dealing with randomly heterogeneous porous media in a high performance computing context, and Jones and Woodward, 2001 [70], Hardelauf et al., 2007 [41] and Herbst et al., 2008 [37] studied the scalability of their Richards Equation solvers for randomly heterogeneous porous media, but the link between scalability and heterogeneity of the considered porous medium still needs to be assessed. Beside the heterogeneity of the medium itself, another kind of heterogeneity may occur during a flow in a variably saturated porous medium: the transition between saturated and unsaturated conditions, which involves strong modifications of the hydrodynamic behaviour. Jones and Woodward [70] discussed the impact on scalability for their implementation (mainly linked to the non-linearity of the soil hydrodynamic parameters) of the coexistence of saturated areas and unsaturated areas. In order to analyse the impact on parallel performances of a Picard/PCG approach of both types of heterogeneity (spatial heterogeneity of the soil hydrodynamic properties and occurrence of transition between saturated and unsaturated zone), three additional scaling curves were performed for a random, spatially uncorrelated heterogeneous (from the point of view of the saturated hydraulic conductivity field) porous medium. The cases concerned were parallelepipedic soil blocks of 20m × 10m × 10m, discretized by regular grids of about 34 millions cells (cell side: about 4cm). The saturated hydraulic conductivity field $K_{sat}$ was a random field with the commonly encountered lognormal distribution (e.g.: [95]), with a mean value of $10^{-6}$m.s$^{-1}$ and a ratio $\sigma(\log(K_{sat}))/\text{mean}(\log(K_{sat}))$ equal to 3.16 ($\sigma$ stands for standard deviation), which is a strong heterogeneity ($K_{sat}$ range : from $10^{-3}$m.s$^{-1}$ to $10^{-10}$m.s$^{-1}$) . The boundary conditions were no-flux boundary conditions on vertical boundaries, with a 5m pressure head at the bottom (presence of a water table at mid-height of the medium). The initial condition is a hydrostatic pressure field, so at the initial time the lower half of the medium is saturated while the upper half is unsaturated. Various regimes of infiltration are modelled, with three different Dirichlet boundary conditions: (i) an imposed water pressure at the top equal to −0.1m (unsaturated infiltration), (ii) another one equal to 0.1m (saturated infiltration) and (iii) a last one equal to 1m (saturated infiltration with a steep moistening front). Thus from (i) to (iii) the modelled moistening fronts are steeper and steeper. The physical time of the modelled infiltrations was 10 days. We considered a precision of $10^{-6}$m for the PCG solver and a precision of $10^{-4}$m for the Picard loop. The initial time step was 0.1s and the maximum time step was 1 day (never reached). One can see the parallel efficiency from 16 cores (reference) to 512 cores, as well as the total computation times of each calculus in Figure B1.

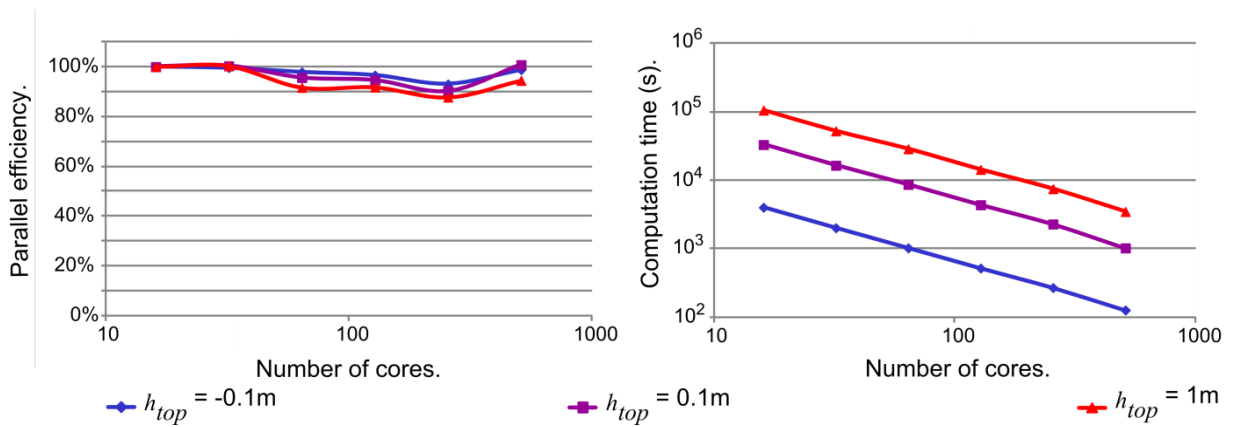

**Figure B1: Parallel efficiencies (left) and computation times (right) from 16 to 512 cores for three different infiltrations with moistening front of increasing steepness in a strongly heterogeneous porous medium (random spatially uncorrelated saturated hydraulic conductivity field).**

The parallel performances are not impacted by the heterogeneity of the porous medium in the case of an infiltration with a smooth moistening front (top water pressure equal to −0.1m). A slight decrease of the parallel efficiency may be observed in the case of a steep front (top water pressure equal to 1m). That may be explained by the fact that in the sub-domains in which the front is located, the resolution of the linear systems generated by the Picard loop requires more iterations of the linear solver PCG that in the sub-domains in which no front are present. It may result in a desynchronisation of the computation times between cores which could explain the observed decrease of parallel efficiency. However, one can note that there is no massive drop of performances. Indeed, the computation times strongly increase with the degree of steepness of the considered infiltration, since more iterations of the Picard loop and of the linear solver PCG is needed when the case get steeper. This is shown, for example, by the evolution with the degree of steepness of the total number of PCG iterations for the whole computation: with a top water pressure equal to -0.1m, we have approximately 3000 iterations, while with a top water pressure equal to 1m we have approximately 233000 iterations. There is thus an increase of almost a factor 80 of the total number of PCG iterations in the steepest case compared to the smoothest. But, although the serial computation time is strongly affected by the steepness of the problem, the impact of this steepness on scalabilty is weak.

Since the scalability of RichardsFOAM seems not to be affected by a spatially uncorrelated heterogeneity of the porous medium, and only slightly by the steepness of the considered problem, we are going to continue the study of the scaling properties of RichardsFOAM by considering results obtained with homogeneous cases and smooth infiltration fronts.

# References

[1] Y. Goddéris, S.L. Brantley, L.M. François, J. Schott, D. Pollard, M. Déqué. Rates of consumption of atmospheric CO2 through the weathering of loess during the next 100 yr of climate change. Biogeosciences Discuss. 9 (2012) 10847

[2] J.C.G. Walker, P.B. Hays, J.F. Kasting. A negative feedback mechanism for the long term stabilization of Earth's surface temperature, J. Geophys. Res. 86 (1981) 9776

[3] Q. Li, A.J.A. Unger, E.A. Sudicky, D. Kassenaar, E.J. Wexler, S. Shikaze. Simulating the multi-seasonal response of a large-scale watershed with a 3D physically-based hydrologic model. J. of Hydrol. 357 (2008) 317

[4] G.-T. Yeh, D.-S. Shih, J.-R.C. Cheng. An integrated media, integrated processes watershed model. Comput. Fluids 45 (2011) 2

[5] C.T. Miller, C.N. Dawson, M.W. Farthing, T.Y. Hou, J. Huang, C. Kees, C.T. Kelley, H.P. Langtangen. Numerical simulation of water resources problems: Models, methods, and trends. Adv. in Water Resour. 51 (2013) 405

[6] L.A. Richards L.A.. Capillary conduction of liquids through porous medium. Physics 1 (1931) 318

[7] M. Muskat. 1949. Physical Principles of Oil Production. McGraw-Hill, New York (1949) p. 922

[8] S. Whitaker. Flow in porous media II: The governing equations for immiscible two-phase flow. Transport Porous Med. 1 (1986) 105

[9] M. Quintard, S. Whitaker. Two-phase flow in heterogeneous porous media I: the influence of large spatial and temporal gradients. Transport Porous Med. 5 (1990) 341

[10] S.M. Hassanizadeh & W.G. Gray. 1997. Recent Advances in Theories of Two-Phase Flow in Porous Media. In Fluid Transport in Porous Media, ed. J.P.D. Plessis (Computational Mechanics Publications, 1997) p. 105

[11] R. Hilfer. Macroscopic equations of motion for two-phase flow in porous media. Phys. Rev. E 58 (1998) 2090

[12] M. Panfilov, I. Panfilova. Phenomenological meniscus model for two-phase flows in porous media. Transport porous med. 58 (2005) 87

[13] L. Cueto-Felgueroso, R. Juanes. A phase field model of unsaturated flow. Water Resour. Res. 45 (2009) W10409. http://dx.doi.org/10.1029/2009WR007945.

[14] W. Wang, J. Rutqvist, U.-J. Görke, J.T. Birkholzer, O. Kolditz. Non-isothermal flow in low permeable porous media: a comparison of Richards' and two-phase flow approaches. Environ. Earth Sci. 62, (2011) 1197

[15] M.-S. Chiang, H.-S. Chu. Numerical investigation of transport component design effect on a proton exchange membrane fuell cell. J. Power Sources 160 (2006) 340

[16] S.A. Galindo-Torres, A. Scheuermann, L. Li, D.M. Pedroso, D.J. Williams. A lattice Boltzmann model for studying transient effects during imbibition-drainage cycles in unsaturated soils. Comput. Phys. Commun. 184 (2013) 1086

[17] R. Ababou & A.C. Bagtzoglou. BIGFLOW: A Numerical Code for Simulating Flow in Variably Saturated, Heterogeneous Geologic Media (Theory and User's Manual, Version 1.1). Report NUREG/CR-6028, U.S. Nuclear Regulatory Commission (Government Printing Office, Washington D.C., U.S.A., 1993) http://www.osti.gov/bridge/servlets/purl/10168217-yoTsuT/10168217.pdf

[18] G. Gottardi, M. Venutelli. Richards: Computer program for the numerical simulation of one-dimensional infiltration into unsaturated soil. Comput. Geosci. 19(9) (1993) 1239

[19] R.H. Brooks, A.T. Corey. Hydraulic Properties of Porous Media. Hydrology Papers, Colorado State University, Fort Collins, Hydrology Paper No.3, March 1964.

[20] M.Th. van Genuchten. A Closed-form Equation for Predicting the Hydraulic Conductivity of Unsaturated Soils. SSSAJ. 44(5) (1980) 892

[21] Y. Mualem. A new model for predicting the hydraulic conductivity of unsaturated porous media. Water Resour. Res. 12 (3) (1976) 513

[22] J. Santos, Y. Efendiev, L. Guarracino. Hydraulic conductivity estimation in partially saturated soils using the adjoint method. Comput. Methods Appl. Mech. Engrg. 196 (2006) 161

[23] M. Bakker, M.W. Farthing, C.S. Woodward. Editorial: Computational challenges in the solution of water resources problems. Adv. Water Resour. 34 (2011) 1059

[24] D.J. Polmann, D. McLaughlin, L.W. Gelhar, R. Ababou. Stochastic modeling of large-scale flow in heterogeneous unsaturated soils. Water Resour. Res., 27(7) (1991) 1447

[25] R. Ababou, S. Sagar, G. Wittmeyer. Testing procedure for spatially distributed flow models, Adv. Water Resour. 15 (1992) 181

[26] C. Paniconi, E.F. Wood. A detailed model for simulation of catchment scale subsurface hydrologic processes. Water Res. Res. 29(6) (1993) 1601

[27] J. Šimůnek, K. Huang, M.Th. van Genuchten. The HYDRUS Code for Simulating the One-Dimensional Movement of Water, Heat, and Multiple Solutes in Variably-Saturated Media. Version 6.0, Research Report No. 144, U.S. Salinity Laboratory, USDA (ARS, Riverside, California, 1998)

[28] J. Šimůnek, M.Th. van Genuchten, M. Šejna. Development and applications of the HYDRUS and STANMOD software packages and related codes. Vadose Zone J. 7(2) (2008) 587

[29] K.U. Mayer, E.O. Frind, D.W. Blowes. Multicomponent reactive transport modeling in variably saturated porous media using a generalized formulation for kinetically controlled reactions. Water Resour. Res. 38(9) (2002) 13-1

[30] S. Weill, E. Mouche, J. Patin. A generalized Richards equation for surface/subsurface flow modelling. J. Hydrol. 366 (2009) 9

[31] M. Camporese, C. Paniconi, M. Putti, S. Orlandini. Surface-subsurface flow modeling with path-based runoff routing, boundary condition-based coupling, and assimilation of multisource observation data. Water Resour. Res. 46(2) (2010) doi: 10.1029/2008WR007536

[32] M. Kuznetsov, A. Yakirevich, Y. Pachepsky Y.A., S. Sorek, N. Weisbrod. Quasi 3D modeling of water flow in vadose zone groundwater, J. Hydrol 450-451 (2012) 140

[33] I. Borsi, R. Rossetto, C. Schifani, M. Hill. Modeling unsaturated zone flow and runoff processes by integrating MODFLOW-LGR and VSF, and creating the new CFL Package. J. Hydrol (2013) doi: http://dx.doi.org/10.1016/j.jhydrol.2013.02.020.

[34] R. Ababou, L.W. Gelhar, C. Hempel. Serial and parallel performance on large matrix. Systems. Cray channels, summer 1992 issue: 22-25 (1992)

[35] F.T. Tracy. Accuracy and Performance Testing of Three-Dimensional Unsaturated Flow Finite Element Groundwater Programs on the Cray XT3 Using Analytical Solutions. HPCMP Users Group Conference (HPCMP-UGC'06) (2006).

[36] D. Coumou, S. Matthäi, S. Geiger, T. Driesner. A parallel FE-FV scheme to solve fluid flow in complex geologic media. Comput. Geosci. 34 (2008) 1697

[37] M. Herbst, S. Gottschalk, M. Reißel, H. Hardelauf, R. Kasteel, M. Javaux, J. Vanderborght, H. Vereecken. On preconditioning for a parallel solution of the Richards equation. Comput. Geosci. 34 (2008) 1958

[38] W. Wang, G. Kosakowski, O. Kolditz. A parallel finite element scheme for thermo-hydro-mechanical (THM) coupled problems in porous media. Comput. Geosci. 35 (2009) 1631

[39] G. Tang, E.F. D'Avezedo, F. Zhang, J.C. Parker, D.B. Watson, P.M. Jardine. Application of a hybrid MPI/OpenMP approach for parallel groundwater model calibration using multi-core computers. Comput. Geosci. 36 (2010) 1451

[40] L. Warsta, T. Karvonen, H. Koivusalo, M. Paasonen-Kivekäs, A. Taskinen. Simulation of water balance in a clayey, subsurface drained agricultural field with three dimensional FLUSH model. J. Hydrol 476 (2013) 395

[41] H. Hardelauf, M. Javaux, M. Herbst, S. Gottschalk, R. Kasteel, J. Vanderborght, H. Vereecken. PARSWMS: A parallelized model for simulating three-dimensional water flow and solute transport in variably saturated soils. Vadose Zone J. 6(2) (2007) 255

[42] K. Zhang, Y. Wu, K. Pruess. User's Guide for TOUGH2-MP – A Massively Parallel Version of the TOUGH2 code. LBNL-315E. Lawrence Berkeley National Laboratory, Berkeley, CA, USA (2008)

[43] G.E. Hammond, P.C. Lichtner, M.L. Rockhold. Stochastic simulation of uranium migration at the Hanford 300 Area. Journal of Contaminant Hydrology 120-121 (2011) 115

[44] M. Williamson, J. Meza, D. Moulton, I. Gorton, M. Freshley, P. Dixon, R. Seitz, C. Steefel, S. Finsterle, S. Hubbard, M. Zhu, K. Gerdes, R. Patterson, Y.T. Collazo. Advanced simulation capability for Environmental Management (ASCEM): an overview of initial results. Tech. Innov., 13(2) (2011) 175

[45] O. Kolditz, S. Bauer, L. Bilke, N. Böttcher, J.O. Delfs, T. Fischer, U.J. Görke, T. Kalbacher, G. Kosakowski, C.I. McDermott, C.H. Park, F. Radu, K. Rink, H. Shao, H.B. Shao, F. Sun, Y.Y. Sun, A.K. Singh, J. Taron, M. Walther, W. Wang, N. Watanabe, Y. Wu, M. Xie, W. Xu, B. Zehner. OpenGeoSys: an open-source initiative for numerical simulation of thermo-hydro-mechanical/chemical (THM/C) processes in porous media. Environ. Earth Sci. 67(2) (2012) 589

[46] C. Lichtner, E. Hammond. Using high performance computing to understand roles of labile and non-labile uranium(VI) on Hanford 300 area plume longevity. Vadose Zone J. (2012) doi: 10.2136/vzj2011.0097.

[47] R.M. Maxwell. A terrain-following grid transform and preconditioner for parallel, large-scale, integrated hydrologic modeling. Adv. Water Resour 53 (2013) 109

[48] H. Jasak. Error Analysis and Estimation for the Finite Volume Method with Applications to Fluid Flows. Ph.D. thesis of the Imperial College of Science, Technology and Medicine, London (1996)

[49] H.G. Weller, G. Tabor, J. Jasak, C. Fureby. A tensorial approach to computational continuum mechanics using object-oriented techniques. Comput. Phys. 12(6) (1998) 620

[50] http://www.openfoam.com

[51] R. Eymard, M. Gutnic, D. Hilhorst. The finite volume method for Richards equation. Computat. Geosci. 3 (1999) 259

[52] I. Rees, I. Masters, A.G. Malan, R.W. Lewis. An edge-based finite volume scheme for saturated-unsaturated groundwater flow. Comput. Methods Appl. Mech. Engrg. 193 (2004) 4741

[53] V. Novaresio, M. García-Camprubí, S. Izquierdo, P. Asinari, N. Fueyo. An open-source library for the numerical modeling of mass-transfer in solid oxide fuel cells. Comput. Phys. Commun. 183 (2012) 125

[54] M. Mortensen, H.P. Langtangen, G.N. Wells. A FeniCS-based programming framework for modeling turbulent flow by the Reynolds-averaged Navier-Stokes equations. Adv. Water Resour. 34 (2011) 1082

[55] J.F. Wellmann, A. Croucher, K. Regenauer-Lieb. Python scripting libraries for subsurface fluid and heat flow simulations with TOUGH2 and SHEMAT. Comput. Geosci. 43 (2012) 197

[56] M.J. Blunt, B. Bijeljic, H. Dong, O. Gharbi, S. Iglauer, P. Mostaghimi, A. Paluszny, C. Pentland. Pore-scale imaging and modelling. Adv. Water Resour. 51 (2013) 197

[57] D.J. Furbish and M.V. Scmeeckle. A probabilistic derivation of the exponential-like distribution of bed-load particle velocities. Water Resour. Res. 49(3) (2013) 1537

[58] I.A. Cosden, J.R. Lukes. A hybrid atomistic-continuum model for fluid flow using LAMMPS and OpenFOAM. Comput. Phys. Commun. 184 (2013) 1958

[59] F. Lehmann, PH Ackerer. Comparison of iterative methods for improved solutions of the fluid flow equation in partially saturated porous media. Transport Porous Med. 31 (1998) 275

[60] H. An, Y. Ichikawa, Y. Tachikawa, M. Shiiba. Comparison between iteration schemes for three-dimensional coordinate-transformed saturated-unsaturated flow model. J. Hydrol. 470-471 (2012) 212

[61] P.A. Lott, H.F. Walker, C.S. Woodward, U.M. Yang. An accelerated Picard method for nonlinear systems related to variably saturated flow. Adv. Water Resour. 38 (2012) 92

[62] G.E. Hammond, A.J. Valocchi, P.C. Lichtner. Application of Jacobian-free Newton-Krylov with physics-based preconditioning to biogeochemical transport. Adv. Water Resour. 28 (2005) 359

[63] C. Paniconi, M. Putti. A comparison of Picard and Newton iteration in the numerical solution of multidimensional variably saturated flow problems. Water Resour. Res. 30(12) (1994) 3357

[64] S. Mehl. Use of Picard and Newton iteration for solving nonlinear ground water flow equations. Groundwater 44(4) (2006) 583

[65] P. Galvao, P. Chambel Leitao, R. Neves, P. Chambel Leitao. A different approach to the modified Picard method for water flow in variably saturated media. Dev. Water Sci. 55 (1) (2004) 557

[66] M.A. Celia, E.T. Bouloutas, R.L. Zarba. A general mass-conservative numerical solution for the unsaturated flow equation. Water Resour. Res. 26 (7) (1990) 1483

[67] C. Paniconi, A.A. Aldama, E.F. Wood. Numerical evaluation of iterative and noniterative methods for the solution of the nonlinear richards equation. Water Resour. Res. 27 (6) (1991) 1147

[68] S.H. Ju, K.-J.S. Kung. Mass types, element orders and solution schemes for the Richards equation. Computers & Geosciences 23(2) (1997) 175

[69] H.J. Lin, D.R. Richards, C.A. Talbot, G.T. Yeh, J.R. Cheng, H.P. Cheng, N.L. Jones. FEMWATER: A three-dimensional Finite Element Computer Model for Simulating Density-Dependent Flow and Transport in Variably Saturated Media". Technical Report CHL-97-12, US Army Engineer Research and Development Center (ERDC), Vicksburg, MS (1997)

[70] J.E. Jones, C.S. Woodward. Newton-Krylov-multigrid solvers for large-scale, highly heterogeneous, variably saturated flow problems. Adv. Water Resour. 24 (2001) 763

[71] T.J.R. Hugues, I. Levit, J. Winget. An element-by-element solution algorithm for problems of structural solid mechanics. Comput. Meth. Appl. Mech. Engr. 36-2 (1983) 241

[72] M.F. Wheeler, M. Peszyńska. Computational engineering and science methodologies for modelling and simulation of subsurface applications. Adv. Water Resour. 25 (2002) 1147

[73] R. Ababou, D. McLaughlin, L.W. Gelhar, A.F.B Tompson. Numerical Simulation of Three-Dimensional Saturated Flow in Randomly Heterogeneous Porous Media. Transport Porous Med. 4 (1989) 549

[74] J.C. van Dam, R.A. Feddees. Numerical simulation of infiltration, evaporation and shallow groundwater levels with the Richards equation. J. Hydrol. 233 (2000) 72

[75] H. Li, M.W. Farthing, C.T. Miller. Adaptive local discontinuous Galerkin approximation to Richards' equation. Adv. Water Resour. 30 (2007) 1883

[76] K. Rathfelder, L.M. Abriola. Mass conservative numerical solutions of the head-based Richards equation. Water Resour. Res. 30 (1994) 3357

[77] K. Kosugi. Comparison of three methods for discretizing the storage term of the Richards equation. Vadose Zone J. 7 (2008) 957

[78] G.A. Williams, C.T. Miller. An evaluation of temporally adaptive transformation approaches for solving Richards' equation. Adv. Water Resour. 22(8) (1999) 831

[79] D. Kavetski, P. Binning, S.W. Sloan. Adaptive time stepping and error control in a mass conservative numerical solution of the mixed form of Richards equation. Adv. Water Res. 24 (2001) 595

[80] K.E. Brenan, S.L. Campbell, L.R. Petzold. The Numerical Solution of Initial Value Problems in Differential-Algebraic Equations. SIAM Frontiers in Applied Mathematics 16 (1996). SIAM, Philadelphia, PA.

[81] B. Belfort, J. Carrayrou, F. Lehmann. Implementation of Richardson extrapolation in an efficient adaptive time stepping method: applications to reactive transport and unsaturated flow in porous media. Transport Porous Med. 69 (2007) 123

[82] S.E. Gasda, M.W. Farthing, C.E. Kees, C.T. Miller. Adaptive split-operator methods for modeling transport phenomena in porous medium systems. Adv. Water Resour. 34 (2011) 1268

[83] J. Carrayrou, J. Hoffmann, P. Knabner, S. Kräutle, C. De Dieuleveult, J. Erhel, J. Van der Lee, V. Lagneau, K. Ulrich, K.T.B. MacQuarrie. Comparison of numerical methods for simulating strongly nonlinear and heterogeneous reactive transport problems – the MoMaS benchmark case. Comput. Geosci. 14 (2010) 483

[84] B. Hendrickson. Load balancing fictions, falsehoods and fallacies. Applied Mathematical Modelling 25(2) (2000) 99

[85] C.E. Kees, M.W. Farthing, C.N. Dawson. Locally conservative, stabilized finite element methods for variably saturated flow. Comput. Methods Appl. Mech. Engrg. 197 (2008) 4610

[86] S.K. Tomer. Soil Moisture Modelling, Retrieval from Microwave Remote Sensing and Assimilation in a Tropical Watershed. Ph.D. thesis, Faculty of Engineering of the Indian Institute of Science, Bangalore (2012)

[87] openfoamwiki.net/index.php/Contrib/groovyBC

[88] Y. Goddéris, L.M. François, A. Probst, J. Schott, D. Moncoulon, D. Labat, D. Viville. Modelling weathering processes at the catchment scale : the WITCH numerical model. Geochimica et Cosmochimica Acta 70 (2006) 1128

[89] J.-J. Braun, M. Descloitres, J. Riotte, S. Fleury, L. Barbiéro, J.-L. Boeglin, A. Violette, E. Lacarce, L. Ruiz, M. Sekhar, M.S. Mohan Kumar, S. Subramanian, B. Dupré. Regolith mass balance inferred from combined mineralogical, geochemical and geophysical studies: Mule Hole gneissic watershed, South India. Geochim. Cosmochim. Ac. 73 (2009) 935

[90] S. Kumar, M. Sekhar, S. Bandyopadhyay. Assimilation of remote sensing and hydrological data using adaptive filtering techniques for watershed modelling. Curr. Sci. India 97(8) (2009) 1196

[91] M. Sekhar, J.-J. Braun, K.V. Hayagreeva Rao, L. Ruiz, H. Robain, J. Viers, J.R. Ndam, B. Dupré. Hydrogeochemical modeling of organo-metallic colloids in the Nsimi experimental watershed, South Cameroun. Environ Geol 54 (2008) 831

[92] M.-L.Bagard, F. Chabaux, O.S. Pokrovsky, J. Viers, A.S. Prokushkin, P. Stille, S. Rihs., A.-D. Schmitt, B. Dupré. Seasonnal variability of element fluxes in two central Siberian rivers draining high latitude permafrost dominated areas. Geochim. Cosmochim. Acta 75 (2011) 3335

[93] W.R. Gardner. Some steady-state solutions of the unsaturated moisture flow equation with application to evaporation from a water table. Soil Sci. 85 (1958) 228

[94] A.C. Bagtzoglou, G.W. Wittmyer, R. Ababou, B. Sagar. Application of a massively parallel computer to flow in a variably saturated heterogeneous porous media. Proceedings of the Computational Methods in Water Resources (1992) 695

[95] G. Martinez, Y. A. Pachepsky, H. Vereecken, H. Hardelauf, M. Herbst, K. Vanderlinden. Modeling local control effects on the temporal stability of soil water content. J Hydrol 481 (2013) 106